\documentclass[aps,pra,reprint,10pt,a4paper]{revtex4-1}
\usepackage[utf8]{inputenc}
\usepackage[sc,osf]{mathpazo}
\usepackage{amsmath}
\usepackage[T1]{fontenc}
\usepackage{latexsym}
\usepackage{amssymb}
\usepackage[colorlinks=true,citecolor=blue,urlcolor=blue]{hyperref}
\usepackage{color}
\usepackage{graphics,epstopdf}
\usepackage{soul}
\usepackage[demo]{graphicx}
\usepackage{capt-of}
\usepackage{lipsum}
\usepackage{adjustbox}
\usepackage[normalem]{ulem}
\usepackage[table,xcdraw]{xcolor}
\usepackage{braket}
\usepackage{physics}
\usepackage{ragged2e}

\usepackage{comment}

\newcommand{\hide}[1]{}

\begin{document}

\title{Sequential information theoretic protocols in continuous variable systems}

\author{Sudipta Das$^1$, Ayan Patra$^2$, Rivu Gupta$^{2,3}$,  Aditi Sen(De)$^2$, Himadri Shekhar Dhar$^{1,4}$}

\affiliation{$^1$ Department of Physics, Indian Institute of Technology Bombay, Powai, Mumbai, Maharashtra - $400076$, India \\
$^2$ Harish-Chandra Research Institute,  A CI of Homi Bhabha National Institute, Chhatnag Road, Jhunsi, Prayagraj - $211019$, India \\
$^3$ Dipartimento di Fisica ``Aldo Pontremoli'', Università degli Studi di Milano, I-$20133$ Milano, Italy\\
$^4$ Centre of Excellence in Quantum Information, Computation, Science and Technology, Indian Institute of Technology Bombay, Mumbai - $400076$, India}

\begin{abstract}
In order to enable the sequential implementation of quantum information theoretic protocols in the continuous variable framework, we propose two schemes for resource reusability, {\it resource-splitting protocol} and {\it unsharp homodyne measurements}. We demonstrate the advantage offered by the first scheme in implementing sequential attempts at continuous variable teleportation when the protocol fails in the previous round.
In the second scheme, unsharp quadrature measurements are employed to implement the detection of entanglement between several pairs of observers. 
Under specific conditions, our calculations show that it is possible to successfully witness the entanglement of the same two-mode state, sequentially by as many as five observers.

\end{abstract}

\maketitle

\section{Introduction}
\label{sec:intro}

The superior efficiency of quantum devices compared to their classical counterparts is primarily attributed to resources that are inaccessible in the classical world.
For example, entanglement~\cite{Horodecki_RMP_2009, Das_QI_2016} is a key resource for quantum communication~\cite{Bennett_PRL_1992, Bennett_PRL_1993}, quantum cryptography~\cite{Ekert91, Gisin_RMP_2002}, and  measurement-based quantum computing~\cite{Briegel_Nature_2009}, while resources such as coherence~\cite{Streltsov_RMP_2017}  are necessary for quantum algorithms~\cite{Deutsch_PRSL_1992, Shor_IEEE_1994, Grover_ACM_1996, Cleve_PRSL_1998}.
Towards realizing these quantum information processing tasks in laboratories, two key types of quantum resources are employed -- on the one hand, 
discrete variable (DV) states having finite degrees of freedom, such as the spin degree of freedom~\cite{Loss_PRA_1998, Kane_Nature_1998, Mills_SA_2022}, polarization of light~\cite{Bennett_JC_1992, Knill_Nature_2001}, and energy levels of an atom~\cite{Henriet_Quantum_2020, Bluvstein_Nature_2022, Graham_Nature_2022, Park_PRX_2022}; on the other hand, continuous variable (CV) states correspond to quantization of electromagnetic fields and canonical position and momentum degrees of freedom~\cite{Ferraro_Bib_2005, Adesso_OSID_2014, Serafini_2017}. The latter choice is motivated by the fact that CV systems can provide some benefits over DV systems, including unconditional preparation of the states required for quantum information processing tasks \cite{Walls-Millburn_1994, Braunstein_RMP_2005}, manipulation of unitaries~\cite{Reck_PRL_1994}, and overcoming the problem of distinguishing entangled measurement bases~\cite{Calsamiglia_APB_2001}. Moreover, they have emerged as the leading candidate for practical demonstrations of quantum supremacy~\cite{Chabaud_PRA_2017, Eisert_Nature_2020, Chabaud_PRR_2021, Chabaud_Quantum_2021, Calcluth_Quantum_2022, Calcluth_PRA_2023, Calcluth_PRX_2024}, and also turn out to be useful for implementing universal quantum gates in quantum circuits~\cite{LLoyd_PRL_1999, Houhou_PRA_2022, Kundra_PRX_2022, Zheng_PRA_2023, McConnell_PRA_2024, Hillman_PRL_2020, Chapman_PRX_2023, Zhou_NPJ_2023, Strandberg_PRL_2024, Eriksson_NC_2024} and building measurement-based quantum computers~\cite{Menicucci_PRL_2006}. 

The implementation of quantum protocols in any physical platform faces several obstacles. A key limitation is the availability of quantum resources, which are not only affected by environmental decoherence \cite{Breuer_2007, Lidar_arXiv_2019, Rivas_2012} but are also readily destroyed during the implementation of a protocol. For instance, projective measurements during a quantum protocol immediately destroy the inherent quantum resource, rendering the post-measurement ensemble useless for any further use. To overcome the limitations due to the latter, innovative techniques to enable the reusability of quantum resources for the sequential implementation of quantum protocols have been developed, often involving the use of \emph{weak} or \emph{unsharp} measurements instead of projective ones~\cite{Silva_PRL_2015}. Such imperfect measurements provide a trade-off between the amount of resource consumed and the proffered quantum advantage. They have been used to demonstrate that non-locality~\cite{Brunner_RMP_2014} can be shared among a maximum of two observers on one side~\cite{Silva_PRL_2015, Mal_Maths_2016}, while the same for entanglement detection extends up to twelve~\cite{Bera_PRA_2018} (see~\cite{Sasmal_PRA_2018, Shenoy_PRA_2019} for the case of quantum steering, and~\cite{Halder_PRA_2022} for bi-nonlocality). Later works have shown how nonlocality and entanglement can be witnessed by an arbitrary number of observer pairs~\cite{Brown_PRL_2020, Pandit_PRA_2022, Srivastava_PRA_2022}, even in the measurement-device-independent~\cite{Srivastava_PRA_2021} and multipartite regimes~\cite{Maity_PRA_2020, Gupta_PRA_2021, Srivastava_arXiv_2022, Srivastava_arXiv_2022_2} (see also~\cite{Mohan_NJP_2019, Miklin_PRR_2020, Curchod_PRA_2017}). 
Sequential protocols have also been proposed for quantum communication schemes such as teleportation~\cite{Roy_PLA_2021} and telecloning~\cite{Das_PRA_2023}. For experimental realizations of such protocols, see Refs.~\cite{Schiavon_QST_2017, Hu_NPJ_2018, Foletto_PRAppl_2020, Choi_Optica_2020, Feng_PRA_2020}.

Although the aforementioned weak or unsharp measurement-based sequential quantum protocols have primarily been implemented using discrete variable systems, the development of sequential protocols for CV systems is quite limited, with the exception of a recent proposal for quantum state estimation via unsharp homodyne measurements, using auxiliary systems~\cite{Das_PRA_2014, Das_JPA_2017}. 
In this work, we propose two strategies that enable the reusability of CV quantum resources to sequentially implement a quantum information theoretic protocol. 
The first protocol, named  {\it resource-splitting}, splits or distributes the resource of the original state among multiple copies using only linear optical elements and resourceless auxiliary states, whereas the second scheme exploits \emph{weak} or \emph{unsharp} measurements of the CV quadratures to allow for resource reusability.



Using CV quantum teleportation~\cite{Braunstein_PRL_1998}, we demonstrate how the resource-splitting scheme can be employed to implement the protocol sequentially multiple times.
We analytically derive the expression of the fidelity in terms of the number of possible rounds and investigate the performance under two constraints, those of equal resource-splitting parameter and equal quantum advantage in each round. While evaluating the maximum number of trials, we establish a relationship between the splitting parameter, the initial squeezing strength of the resource, and the quantum advantage. Interestingly, we exhibit that a substantial number of rounds may be possible if the output fidelity is demanded to just surpass the classical threshold.
To illustrate the effectiveness of the unsharp measurement scheme, we apply it to the sequential detection of entanglement of a two-mode Gaussian state. Specifically, using auxiliary squeezed vacuum states and suitable unitary operators~\cite{Das_PRA_2014, Das_JPA_2017}, we provide an upper bound on the number of parties that can sequentially detect entanglement when all the rounds employ the same unsharp measurement parameter. 
Similar bounds can also be obtained when the pairs of observers choose different unsharp parameters. Depending on the initial squeezing strength, we show that up to five observers can sequentially detect entanglement of the same two-mode state. This demonstrates the complementary relationship between unsharp measurements and quantum advantage in continuous variable setups.

The article is organized in the following manner. In Sec.~\ref{sec:re-splitting}, we introduce the resource-splitting protocol for CV resource states and demonstrate its efficacy in Sec.~\ref{sec:multiple-teleportation} through the CV teleportation protocol. The implementation of the unsharp measurement scheme is described in Sec.~\ref{sec:weak_meas}, and the sequential detection entanglement using the inseparability criterion is presented in Sec.~\ref{sec:seq_ent}. We end our paper with discussions in Sec.~\ref{sec:conclu}.

\section{Reusability of resource via resource-splitting}
\label{sec:re-splitting}

In quantum information theoretic protocols, projective measurements of rank-$1$ are typically performed to extract maximum quantum information~\cite{Bennett_PRL_1992, Bennett_PRL_1993, Braunstein_PRL_1998, Braunstein_PRA_2000, Grosshans_Nature_2003, Curty_PRL_2004}. A direct consequence of such measurements is that they destroy the quantum correlated shared resource, rendering the post-measurement ensemble unusable for any future protocols. 
To overcome this, several schemes have been proposed over the years~\cite{Derka_PRL_1998, Shang_PRA_2018, Dieks_PLA_1988, Peres_PLA_1988, Vertesi_PRA_2010, Gomez_PRA_2018, Halder_PRA_2021, Mondal_arXiv_2023}, primarily using non-projective or unsharp measurements~\cite{nielsenchuang, Serafini_2017}.

In CV systems, we adopt a strategy based on splitting the CV resource to reuse the state. This is achieved by combining each mode with a resourceless auxiliary mode and performing a local unitary operation, thereby creating two imperfect copies of the initial resource. 
We refer to our scheme as the \emph{resource-splitting} method. Importantly, the resource content of the two copies, and thus their ability to offer quantum advantage, can be controlled by the parameters of the auxiliary state and the unitary operator.
\begin{figure*}[ht]
\includegraphics[width=.7\linewidth]{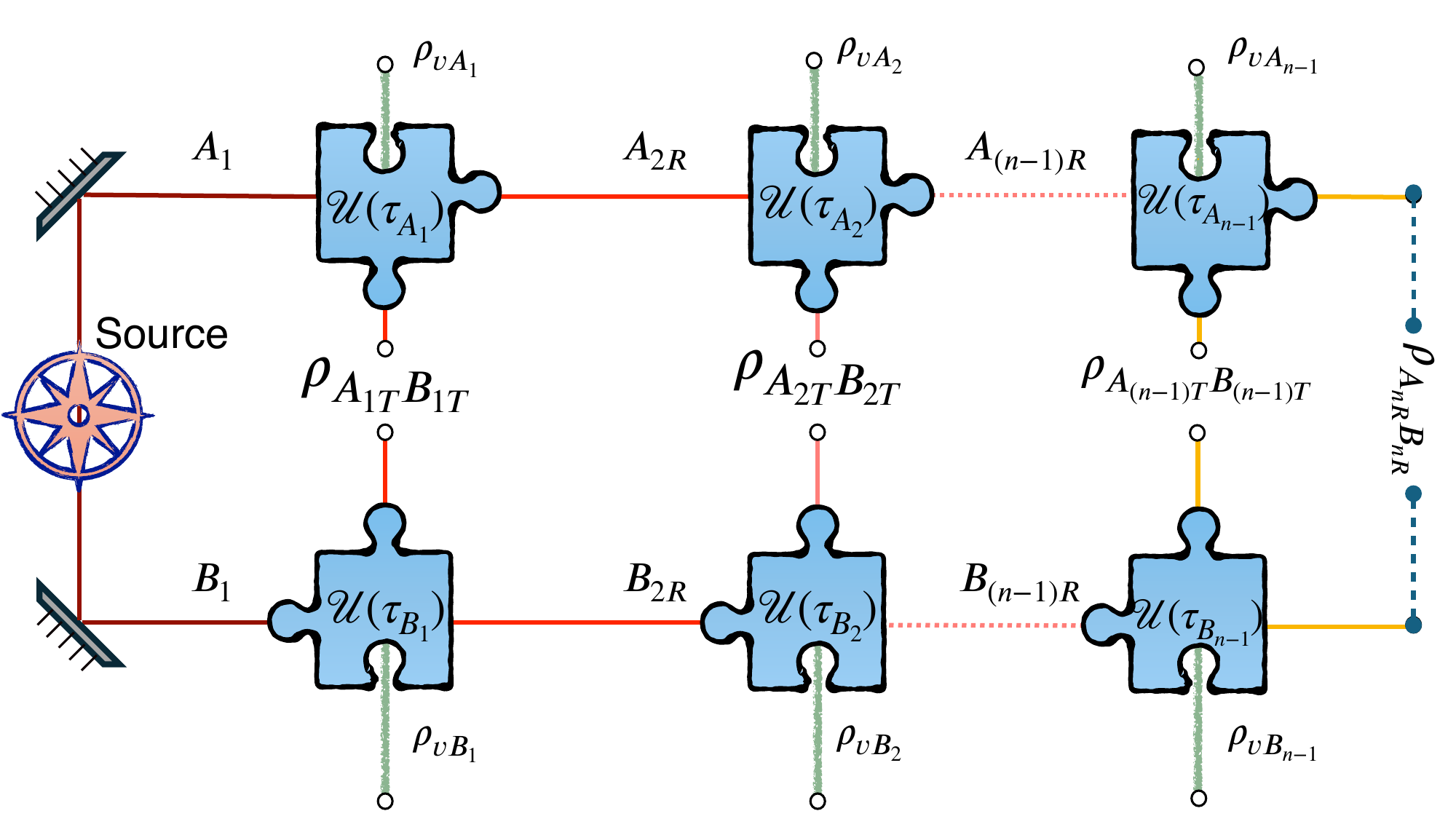}
\caption{{\bf Schematic of the resource-splitting protocol.} Two parties, Alice and Bob, situated in distant locations, share a two-mode entangled state with modes $A_1$ and $B_1$. Both parties combine their modes with respective vacuum states, $\rho_{vA_1}$ and $\rho_{vB_1}$, and apply  the unitary operation,  $\mathcal{U}(\tau_{A_1},\tau_{B_1})=\mathcal{U}(\tau_{A_1})\otimes\mathcal{U}(\tau_{B_1})$, with $\tau_{A_1}$ (\(\tau_{B_1}\)) being the parameters of the unitary operator. Here  $\mathcal{U}(\tau_{A_1})$ $( \mathcal{U}(\tau_{B_1}))$ are beam splitters belonging to Alice (Bob) with transmissivity  $\tau_{A_1}$ (\(\tau_{B_1}\)). The transmitted resource state, $\rho_{A_{1T}B_{1T}}$, can be utilized in the quantum communication protocol, while the reflected resource, denoted as $\rho_{A_{2R}B_{2R}}$ can be further reused. The same process can be continued  \(n-1\) times producing the transmitted and reflected states  $\rho_{A_{(n-1)T}B_{(n-1)T}}$  and $\rho_{A_{nR}B_{nR}}$, respectively.} 
\label{fig:sch1} 
\end{figure*}
To illustrate the protocol, consider two parties, Alice (A) and Bob (B), that share a two-mode entangled state, $\rho_{AB}$, as the resource. Both parties combine their respective modes with an auxiliary vacuum mode, $\rho_{v_i} = |{0}\rangle\langle 0|_i$, using a beam splitter, $\mathcal{U}(\tau_i)$, of transmissivity $\tau_i (i=\text{A, B})$, as shown in Fig.~\ref{fig:sch1}. 
We note that a squeezed vacuum state can also be used as an auxiliary mode.
However, this would result in additional asymmetric noise in the quadratures~\cite{van-Loock_PRL_2001}. 
The composite system after the splitting process can be written as
\begin{eqnarray}
    \rho_{\text{out}}=\mathcal{U}(\tau_A,\tau_B) \rho_{\text{in}}\mathcal{U}^\dagger(\tau_A,\tau_B),
    \label{eq_splitting_res}
\end{eqnarray}
where $\rho_{in}=\rho_{v_A}\otimes\rho_{AB}\otimes\rho_{v_B}$ is the initial state and $\mathcal{U}(\tau_A,\tau_B)=\mathcal{U}(\tau_A)\otimes\mathcal{U}(\tau_B)$ is the total unitary operation on both the parties, with \textcolor{black}{the local beam splitter unitary given as} $\mathcal{U}(\tau_i)=e^{\iota \theta \left( \hat{a}_{i}^{\dagger} \hat{a}_{v_i} + \hat{a}_{i} \hat{a}_{v_i}^{\dagger} \right)}$, where $\hat{a}_{i}$ is the annihilation operator of the mode, $\iota = \sqrt{-1}$, $i \in \{A, B\}$, and $\theta = \cos^{-1} \tau_i$. The state $\rho_{\text{out}}$ comprises four modes - two reflected and two transmitted at each end and can be represented as $\rho_{A_TA_RB_TB_R}$ where $R$ and $T$ depict the reflected and transmitted parts, respectively.

As such, the shared resource $\rho_{AB}$ is now split into a pair of resources -- the transmitted state, $\rho_{A_T B_T}=\Tr_{A_R, B_R} (\rho_{\text{out}})$, and the reflected state, $\rho_{A_R B_R}=\Tr_{A_T, B_T} (\rho_{\text{out}})$. In any quantum information protocol, the transmitted state can then be consumed, while the reflected state serves as the resource for the next round of the protocol. It is expected that resource-splitting effectively redistributes the resource content among the split modes, which leads to a tradeoff between optimal performance and the reusability of the state during the protocol. In other words, reusability comes at the cost of high quantum advantage.

\section{Sequential teleportation using reusable resource}
\label{sec:multiple-teleportation}

Let us demonstrate the usefulness of the resource-splitting scheme in a communication protocol, namely CV teleportation~\cite{Braunstein_PRL_1998}. 
Consider the situation where the classical communication step involved in the teleportation protocol fails, or the receiver, Bob ($B$), is unable to complete the protocol after the post-measurement results have been classically communicated to him by the sender, Alice ($A$).
In the ideal case, Alice performs a projective measurement, and any failure on Bob's part negates the possibility of completing the protocol unless the entire process is repeated with a new resource state.
In discrete variable systems, protocols with multiple attempts to teleport quantum states to Bob have been designed, where the sender, Alice, performs optimal positive-operator valued measurements (POVMs)~\cite{Roy_PLA_2021, Das_PRA_2023}, instead of projective Bell measurements. 
In CV systems, we show that the resource-splitting protocol is an effective approach to create reusable resources for sequential attempts at quantum teleportation.
Note that since the receiver is deemed unable to complete the teleportation protocol, for this discussion, it is sufficient to consider splitting the resource only at the sender's end. 

Suppose Alice and Bob share a two-mode squeezed vacuum (TMSV) state~\cite{Einstein_PR_1935}. Alice combines her mode with a vacuum state, $\ketbra{0}{0}_A$, using a beam splitter of transmissivity, $\tau_1$, to split the resource at her end. She
then performs a double-homodyne measurement on the transmitted mode and the single-mode input coherent state to be teleported. 
The fidelity of the teleported state, if Bob successfully completes the protocol, is given by
\begin{eqnarray}
    \mathcal{F}_1=\frac{2}{3-\tau_1+(1+\tau_1)\cosh{2r}-2\sqrt{\tau_1}\sinh{2r}},
    \label{eq:first_round}
\end{eqnarray}
where $r$ is the squeezing strength of the original TMSV state. A detailed derivation of the expression for the fidelity is given in Appendix~\ref{app:recycle_protocol}.
Note that the fidelity $\mathcal{F}_1$ does not depend on the displacement parameter of the input coherent state, which
allows us to interpret it
as the average fidelity. Substituting  $\tau_1=1$ into Eq.~\eqref{eq:first_round}, the fidelity corresponding to ideal teleportation, $\mathcal{F}_{\text{ideal}} = 1/(1 + e^{-2r})$, can be retrieved. 
Importantly, in case the protocol fails in the first round, Alice still shares a two-mode resource state with Bob, consisting of the reflected part of her split resource and Bob's unused mode. This allows Alice to attempt the teleportation protocol again.
Alice splits the new resource using another auxiliary vacuum state and a beam splitter with transmissivity $\tau_2$ and once more employs the transmitted mode to teleport the coherent state. 
If the resource splitting scheme is sequentially performed multiple times, the  average fidelity of teleportation at some round, $n$, when the protocol finally succeeds, turns out to be
\begin{eqnarray}
    \mathcal{F}_n=\frac{2}{3-\tau^{(n)}_t+(1+\tau^{(n)}_t)\cosh{2r}-2\sqrt{\tau^{(n)}_t}\sinh{2r}}
    \label{eq:genfid_0},
\end{eqnarray}
where $\tau^{(n)}_t=(1-\tau_1)(1-\tau_2).....(1-\tau_{n-1})\tau_n$ and $\tau_i$ denotes the transmissivity in the round $i$, with the superscript $(n)$ denoting the round number. 
Therefore, the sender, Alice's, local operation can repeat the process in such a way that all $n$ rounds can be characterized by $n + 1$ parameters, $(r, \tau_1, \tau_2, ..., \tau_n)$ in the fidelity expression.
\begin{figure*}[ht]
\includegraphics[width=.7\linewidth]{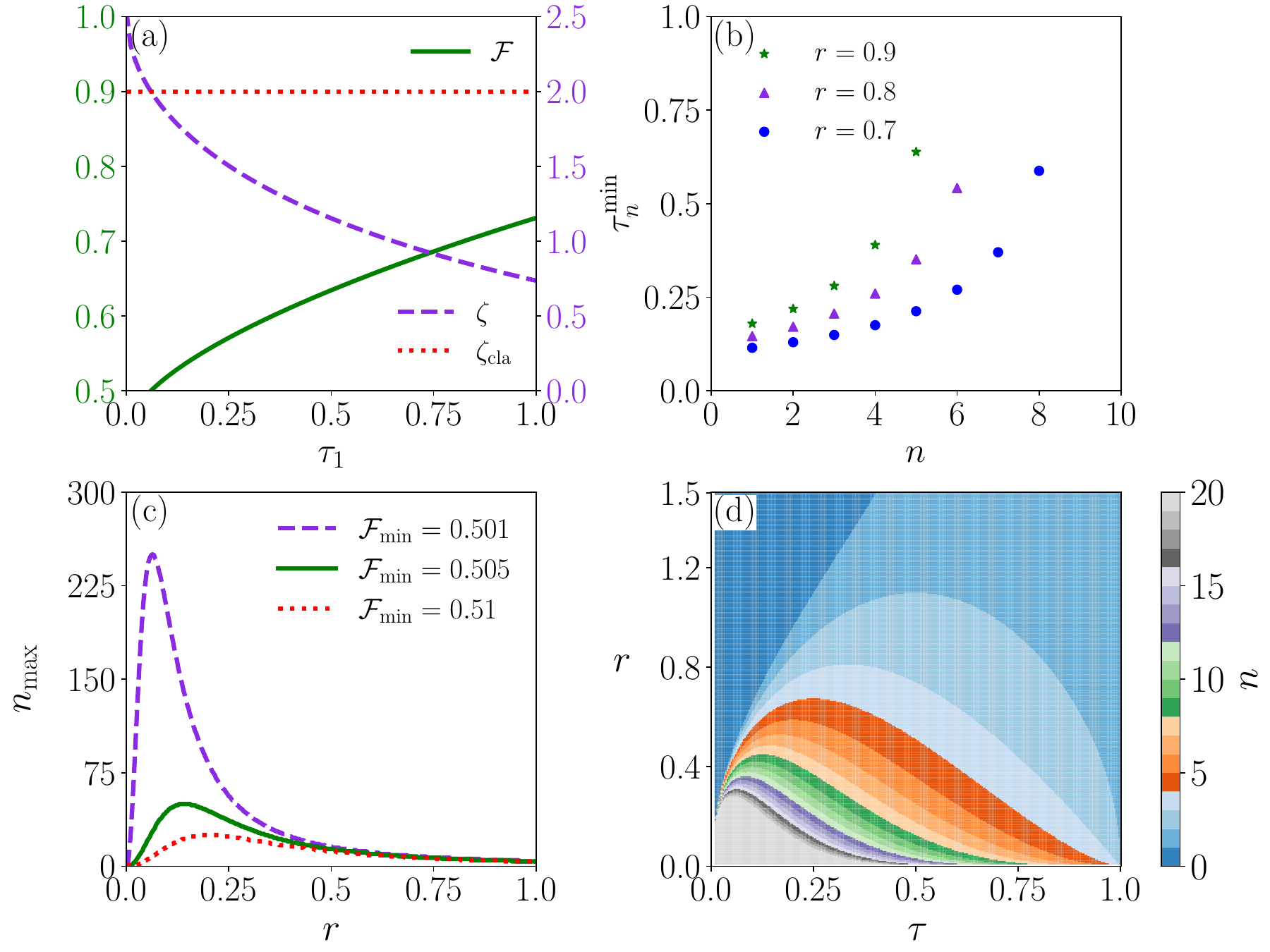}
\caption{\textbf{Use of  resource-splitting protocol in CV teleportation}. (a) Solid (green) and dashed (blue-violet) lines represent the fidelity, $\mathcal{F}$ (left ordinate), and the non-classicality measure, $\zeta$ (right ordinate), respectively, with respect to the transmissivity, \(\tau_1\) (abscissa) of the beam-splitter for the first round. The classical fidelity, $\zeta_{\text{cla}} = 2$ (dotted (red) line), is plotted with respect to \(\tau_1\) for reference. (b) Plot of the minimum transmissivity of round $n$, i.e., $\tau_n^{\min}$ (ordinate) against the round number, $n$, (abscissa). Here we consider the squeezing strength of the shared resource state as $r=0.9$ (green star), $r=0.8$ (blue-violet triangle) and $r=0.7$ (blue circle), by fixing the transmissivity of all proceeding rounds to be at $\tau_i^{\min}\; \forall \; i < n$. (c) The maximum round number, $n_{\max}$ (ordinate) versus the squeezing strength $r$ (abscissa) for the minimum equal fidelity  $\mathcal{F}_{\text{min}}=0.501$ (blue-violet dashed line), $\mathcal{F}_{\text{min}}=0.505$ (green solid line), $\mathcal{F}_{\text{min}}=0.51$ (red dotted line), respectively. (d) Color map of the round number $n$ against the squeezing strength, $r$ (ordinate), and transmissivity, $\tau$ (abscissa), for {equal transmissivity} configurations in every round. All axes are dimensionless.}
\label{fig:fid_tin0} 
\end{figure*}

Before investigating the sequential teleportation scheme further, let us recall the non-classicality measure previously defined in~\cite{Duan_PRL_2000}
 \begin{eqnarray}
     \zeta(t,b)=\langle \Delta^2 (\hat{x}_t - \hat{x}_b) \rangle + \langle \Delta^2 (\hat{p}_t + \hat{p}_b) \rangle,
     \label{eq:q_fn}
 \end{eqnarray}
where $t$ denotes the transmitted mode of the split resource with Alice, and $b$ is the mode with the receiver Bob.
It can be shown that quantum fidelity at each round, i.e., $\mathcal{F} > 0.5$, is achieved when $0 \leq \zeta({t, b}) < \zeta_{\text{cla}} = 2$~\cite{Das_PRA_2024}.
For round $n$, the non-classicality measure is given by
\begin{eqnarray}
    \nonumber \zeta_{n}(t,b)&=&2-2\sinh{r}\big(2\sqrt{\tau_t^{(n)}}\cosh{r}\\
    &-&(1+\tau_t^{(n)})\sinh{r}\big).
    \label{eq:genq_fn}
\end{eqnarray}
Let us describe the sequential teleportation protocol with an initial TMSV state at a fixed squeezing strength, $r = 0.8$. 
In the first attempt, i.e., $n=1$, we observe that the transmitted resource is nonclassical upon splitting, i.e., $\zeta_1(t, b)< 2$, when the transmissivity is above a threshold, $\tau_1 \gtrsim 0.144$.
Therefore, the teleportation fidelity $\mathcal{F}_1$ demonstrates quantum advantage beyond this threshold transmissivity as depicted \textcolor{black}{in Fig.~\ref{fig:fid_tin0}(a).} 
Similarly,
there exists a threshold transmissivity, denoted as $\tau_n^{\min}$, for each round $n$ such that $\zeta_n(t, b)< 2$. 
Now, for a fixed $n$ and $r$, one can calculate $\tau_1^{\min}$ such that the fidelity is just above the classical threshold, i.e., $\mathcal{F}_1 = 0.501 > 0.5$. 
Sequentially, with this value of $\tau_1^{\min}$ in Eq.~\eqref{eq:genfid_0}, one can calculate $\tau_2^{\min}$ \textcolor{black}{to achieve fidelity} just breaching the classical limit, i.e., $\mathcal{F}_2 = 0.501 > 0.5$ and so on.
It is evident \textcolor{black}{from Fig.~\ref{fig:fid_tin0}(b)} that $\tau_n^{\min}$ increases with the increase of $n$, since as $n$ increases, the entanglement of the split resource decreases, and a greater portion of the resource must be transmitted to further extend the sequential protocol.
Ultimately, $\tau_n^{\min}$ cannot be greater than unity, which imposes a restriction on the number of rounds, $n$, for which the protocol can be sequentially implemented.


\subsection{Quantum advantage under different constraints}

The benefit of the proposed resource-splitting scheme in teleportation is emphasized under two constraints in the protocol, viz., the requirement of \textit{equal fidelity} and \textit{equal transmissivity} in all rounds. 
In the first case, we claim that an equal non-classical fidelity, $\mathcal{F}=\mathcal{F}_{\min} > 1/2$, be achieved at all rounds of the protocol, while the latter focuses on the quantum advantage when beam splitters with identical transmissivity, $\tau$, are used to split the resource in every round. In both cases, we evaluate the maximum number of rounds, $n_{\max}$, where resource-splitting offers quantum advantage in sequential teleportation.


\emph{(i) Equal fidelity.--} In this protocol, the sender, Alice, can configure the beam splitter transmissivity in each round such that a fixed, minimum quantum fidelity $\mathcal{F}_{\min}$ is achieved in every round, i.e., $\mathcal{F}_1=\mathcal{F}_2=.......=\mathcal{F}_{n-1}=\mathcal{F}_n = \mathcal{F}_{\min} > 1/2$.  
From Eq.~\eqref{eq:genfid_0}, and the subsequent relation between $\tau_1$, $\mathcal{F}_1$  and $n$, a relation between $\tau_i$ and $\tau_{i-1}$ emerges for a fixed $n$,
\begin{eqnarray}
    \tau_i=\frac{\tau_{i-1}}{1-\tau_{i-1}}.
    \label{eq:recursion}
\end{eqnarray}
Taking $\tau_n=1$ in the final round, $n$, where we assume perfect transmissivity and teleportation,
Eq.~\eqref{eq:recursion} leads to $\tau_1=1/n$. Substituting $\tau_1$ into Eq.~\eqref{eq:first_round}, the fidelity at every round for a fixed $n$ can be expressed as
 \begin{eqnarray}
    \mathcal{F}=\frac{2n}{3n-1+(1+n)\cosh{2r}-2\sqrt{n} \sinh{2r}}.
    \label{eq:eqfid_tin0}
\end{eqnarray}
Therefore, for a constant $\mathcal{F}$, $n$ can be expressed as a function of $r$. We observe that for a fixed $\mathcal{F} = \mathcal{F}_{\min}$, there is an optimal squeezing, $r_{\text{opt}}$, for which $n$ is maximum, as \textcolor{black}{shown in Fig.~\ref{fig:fid_tin0}(c).} The maximum number of rounds of sequential teleportation, $n_{\max}$, can be found to be $n_{\max}={\mathcal{F}}/({2\mathcal{F}-1})$, by expressing Eq.~\eqref{eq:eqfid_tin0} as a function of $r$ and $\mathcal{F}$, and taking derivatives with respect to $r$. The corresponding optimal squeezing is given as  
\begin{eqnarray}
    r_{\text{opt}} = \frac{1}{2}\ln{\left[\frac{\sqrt{\mathcal{F}}+\sqrt{2\mathcal{F}-1}}{\sqrt{\mathcal{F}}-\sqrt{2\mathcal{F}-1}}\right]}.
    \label{r_opt}
\end{eqnarray}
It is observed that as $\mathcal{F} = \mathcal{F}_{\min}$ approaches the classical threshold of $0.5$, $n_{\max}$ can take arbitrarily large, but finite values, and the corresponding optimum squeezing asymptotically vanishes, i.e, $r_{\text{opt}} \rightarrow 0 ~ \text{as}~ \mathcal{F}_{\min} \to 0.5$. When $\mathcal{F}_{\min} = 1/2$, we can have an unbounded number of attempts, but since this denotes the absence of any quantum advantage, the situation is of no practical relevance. 
An interesting observation here is that the optimal $r$ in Eq.~(\ref{r_opt}) does not correspond to the case of a maximally entangled state or $r \rightarrow \infty$. This is primarily due to the fact that the resource-splitting protocol creates asymmetric resources in each round that are shared between unequal numbers of sender and receiver modes. As such, the optimal resource for the specific protocol does not necessarily correspond to infinite squeezing. Similar optimality conditions are also observed in multiparty CV communication protocols such as telecloning~\cite{van-Loock_PRL_2001,Das_PRA_2024}. On the other hand, protocols such as teleportation~\cite{Braunstein_RMP_2005} between a single sender and a single receiver, sharing a symmetric resource, 
have optimal performance when the resource is infinitely squeezed.


Note that the protocols achieving an unbounded number of attempts in the violation of Bell inequalities~\cite{Brown_PRL_2020, Cheng_PRA_2022, Kumari_PRA_2023}, and detection of entanglement~\cite{Pandit_PRA_2022, Srivastava_PRA_2022, Srivastava_arXiv_2022, Srivastava_arXiv_2022_2} in finite-dimensional systems are qualitatively different from the one presented here.



\emph{(ii) Equal transmissivity.--} Consider the protocol where the sender, Alice, uses beam splitters of equal transmissivity, $\tau$, at each round. In such a case, the expression for $\tau_t^{(n)}$ simply reduces to $\tau^{(n)} = (1-\tau)^{n-1}\tau$, where $\tau_i = \tau$ $ \forall i \in \{1,n\}$, and the fidelity of teleportation for round $n$ is given by
\begin{eqnarray}
    \nonumber \mathcal{F}_n&=&\frac{2}{3-\tau^{(n)}+(1+\tau^{(n)})\cosh{2r}-2\sqrt{\tau^{(n)}}\sinh{2r}}.\\
    \label{eq_equal_transmissivity}
\end{eqnarray}
Considering the fidelity of all previous rounds to be greater than the classical benchmark, the condition $\mathcal{F}_n\rightarrow 0.5^{+}$ can determine the maximum round number $n_{\max}$. \textcolor{black}{Fig.~\ref{fig:fid_tin0}(d) indicates} how the attempt number $n$ changes as a function of $r$ and $\tau$. For a fixed $r$, it is observed that there is an optimal transmissivity, $\tau_{\text{opt}}$, for which $n_{\max}$ reaches its highest value. There is a decreasing trend of  $\tau_{\text{opt}}$ as $r$ decreases.

The sequential teleportation protocol under the equal fidelity constraint always dominates over the one with equal transmissivity, particularly in terms of the maximum number of attempts that yield quantum advantage. This happens because the constraint of equal fidelity optimizes the \textit{resource-splitting} scheme.

\section{Resource reusability based on unsharp measurement}
\label{sec:weak_meas}
The second strategy to reuse CV resources in a quantum information protocol is based on weak or unsharp measurements.
Typically, ``weak'' measurements offer partial information about a quantum system while disturbing it minimally~\cite{Derka_PRL_1998, Shang_PRA_2018, Dieks_PLA_1988, Peres_PLA_1988, Vertesi_PRA_2010, Gomez_PRA_2018, Halder_PRA_2021, Mondal_arXiv_2023}, as compared to ``strong'' or projective measurements that completely destroy any available resource at the cost of yielding complete information. 
As such, one of the main motivations for studying unsharp measurement is to
allow resources to be reused, especially in sequential quantum information protocols, apart from the usual scenario of the measurement apparatus being inherently noisy.
For instance, in discrete variable systems, weak or unsharp measurements have been used to sequentially detect Bell violation over multiple observer pairs~\cite{Silva_PRL_2015, Mal_Maths_2016, Bera_PRA_2018}, to witness the entanglement of a state~\cite{Pandit_PRA_2022, Srivastava_PRA_2021, Srivastava_PRA_2022}, for randomness extraction~\cite{Curchod_PRA_2017, Foletto_PRA_2021}, precision metrology~\cite{Zhang_PRL_2015}, and to undertake teleportation~\cite{Roy_PRA_2021}, telecloning~\cite{Das_PRA_2023}, as well as cryptographic schemes~\cite{Troupe_arXiv_2017}. Such measurements have also been employed for state estimation in CV systems~\cite{Das_PRA_2014, Das_JPA_2017}. In this regime, the most well-known and easily implementable projective measurements correspond to homodyne detection of the quadratures~\cite{Ferraro_Bib_2005, Adesso_OSID_2014, Serafini_2017}. Specifically, homodyne measurement is equivalent to measuring 
$|{\hat{x}_\theta}\rangle\langle{\hat{x}_\theta}|$,
where
$\hat{x}_\theta=\cos{\theta}\hat{q}+\sin{\theta}\hat{p}$,
represents a rotated quadrature operator~\cite{Serafini_2017}. 

To highlight the efficacy of resource reusability using unsharp measurements, we demonstrate a protocol of unsharp measurement of the position and momentum quadrature of a two-mode Gaussian entangled state.
\begin{figure*}[ht]
\includegraphics[width=.8\linewidth]{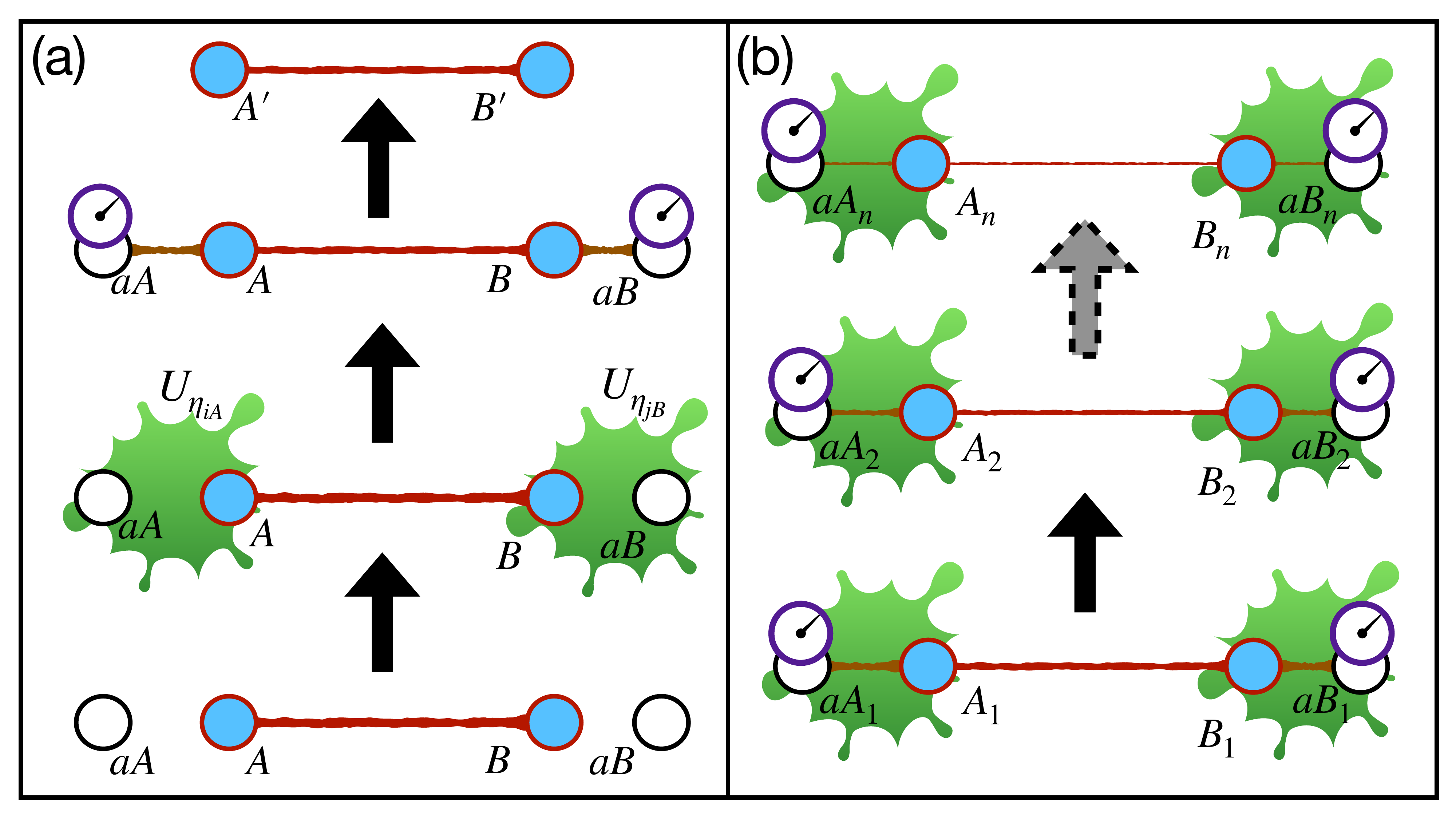}
\caption{\textbf{Schematic of the sequential unsharp measurement scheme.}  (a) {\it Basic set-up.} Two parties, \(A\) and \(B\), share a two-mode entangled state. To measure the quadratures of the shared resource, i.e., $(\eta_{iA},\eta_{jB})$, both parties combine their respective modes with auxiliary modes, $aA$ and $aB$, respectively, and evolve the system using unitary operators, $U_{\eta_{iA},\eta_{jB}}=U_{\eta_{iA}}\otimes U_{\eta_{jB}}$. Measuring the auxiliary modes using projective measurement provides partial information about the quadratures of the initial resource.  (b) \textit{Sequential unsharp measurement protocol.} Two parties, $A_1$ and $B_1$,  unsharply measure their quadratures using the unsharp measurement protocol discussed in (a) and pass the post-measurement resource to another pair of parties, $(A_2, B_2)$. They also independently measure their modes with the help of their respective auxiliary systems, \(aA_2\) and \(aB_2\), before sharing the post-measurement state with the next pair, and so on. The measurements  continue until the entanglement present in the shared post-measurement state disappears.}
\label{fig:weak_measurement} 
\end{figure*}
Consider a two-mode entangled state, initially shared between parties $A$, and $B$. 
The non-classicality between the modes can be measured using $\zeta(A, B)$, as defined in Eq.~\eqref{eq:q_fn}. 
This also happens to be a valid inseparability criterion~\cite{Duan_PRL_2000}, and acts as a witness for Gaussian entanglement. 
A Gaussian state is entangled iff $\zeta(A, B) < 2$.
However, for non-Gaussian states, the criterion is only sufficient.
However, $\zeta(A, B)$ serves as a good figure of merit to detect the entanglement of the resource for our unsharp measurement protocol.
To calculate $\zeta(A, B)$, each party combines their respective modes with auxiliary ($a$) squeezed states and evolves their combined system using the local unitary operator $U_{\eta_{iA},\eta_{jB}}= U_{\eta_{iA}}\otimes U_{\eta_{jB}}$, 
where $\eta_{ik}=\{\hat{q}_{k}, \hat{p}_{k}\}$ and $\chi_{ik}=\{\hat{q}_{ak}, \hat{p}_{ak}\}$  are the quadratures of the resource and auxiliary system respectively, where $i = \{1, 2\} \equiv \{\hat{q}, \hat{p}\}$ represents the quadratures of a mode and the index $k$ denotes the parties, i.e., $k=\{A, B\}$.
The unitary governs the unsharp measurement protocol on the resource quadrature, $\{\eta_{iA},\eta_{jB}\}$, as shown in Fig.~\ref{fig:weak_measurement}(a).
The corresponding interaction Hamiltonian between any mode and its corresponding auxiliary subsystem can be written as~\cite{Das_PRA_2014, Das_JPA_2017}
\begin{eqnarray}
    \mathcal{H}_{\eta_{ik}}=\delta(t-t_0)\eta_{ik}\chi_{2k}.
    \label{eq_hamiltonian}
\end{eqnarray}
As such, $U_{\eta_{iA}} = \exp(- \iota \int \mathcal{H}_{\eta_{iA}} dt)$. 
The commutator relation of the system quadratures is given by $[\eta_{ik},\eta_{jk'}]=-\iota(i-j)\delta_{k,k'}$, and a similar expression holds for the auxiliary system. The Hamiltonian shifts the mean value of the position quadrature of the auxiliary state by an amount  $\eta_{ik}$ such that $\chi_{1k}\rightarrow\chi_{1k}+\eta_{ik}$ depending on the choice of $U_{\eta_{ik}}$.
Therefore, if both parties now independently perform homodyne measurements on the position quadrature of the auxiliary mode, i.e., on $\chi_{1k}=\{\hat{q}_{aA}, \hat{q}_{aB}\}$, the results contain partial information of the system quadrature $\eta_{ik}$, depending upon the unitary evolution.
The function $\zeta(A,B)$ can be computed as follows:
\begin{eqnarray}
    \nonumber\langle\Delta^2(\eta_{1A}-\eta_{1B}) \rangle&=&\langle \Delta^2(\chi_{1A}-\chi_{1B})\rangle_{U_{\eta_{1A},\eta_{1B}}},\\
    \langle\Delta^2(\eta_{2A}+\eta_{2B}) \rangle&=&\langle \Delta^2(\chi_{1A}+\chi_{1B})\rangle_{U_{\eta_{2A},\eta_{2B}}}.
    \label{eq_vanloock_cal}
\end{eqnarray}
In the above equation, the left-hand side indicates the variance of the system quadrature, $\eta_{iA}$, when the quadrature, $\chi_{1A}$, of the auxiliary system is measured after evolution through $U_{\eta_{iA},\eta_{jB}}$ (see Appendix \ref{app:sequential} for a detailed derivation of the unsharp measurement protocol). The quadrature variance of the auxiliary system reads

\begin{eqnarray}
    \nonumber  \langle \chi_{1A}^n\otimes \chi_{1B}^{2-n}\rangle_{U_{\eta_{iA},\eta_{jB}}}= \int && \chi_{1A}^n\chi_{1B}^{2-n}P(\chi_{1A},\chi_{1B})_{\eta_{iA},\eta_{jB}} \times \\
    && \prod_k d\chi_{1k},
    \label{eq:van_loock_prob}
\end{eqnarray}
where $n=\{0,1,2\}$ and $\chi_{1k}^0=\mathbb{I}$, where $\mathbb{I}$ is the identity, and $P(\chi_{1A},\chi_{1B})_{\eta_{iA},\eta_{jB}}$ is the probability of finding the measurement outcome $\chi_{1k}=\{\hat{q}_{aA}, \hat{q}_{aB}\}$, corresponding to the system quadratures $(\eta_{iA},\eta_{jB})$. If both parties share a TMSV state, the inseparability criterion is given as 

\begin{eqnarray}
    \zeta(A,B)=2e^{-2r}+\langle\Delta^2 \chi_{1A}\rangle+\langle\Delta^2 \chi_{1B}\rangle,
    \label{eq:one-round}
\end{eqnarray}
where $\langle\Delta^2 \chi_{ik}\rangle$ is the quadrature variance of the auxiliary system, which, being squeezed vacuum modes, saturate the uncertainty relation, $\langle\Delta^2 \chi_{1k}\rangle\langle\Delta^2 \chi_{2k}\rangle=\frac{1}{4}$. The entanglement of the shared resource $\rho_{AB}$ can be detected using measurements on the auxiliary subsystem if $\zeta(A, B) < 2$. 
This happens when
\begin{eqnarray}
    \langle\Delta^2 \chi_{1,A}\rangle+\langle\Delta^2 \chi_{1,B}\rangle<2(1-e^{-2r}).
\end{eqnarray}
Therefore, the quadrature variance of the auxiliary system works as the unsharp parameter of the measurement. For example, if we consider equal unsharp parameters, i.e., $\langle\Delta^2 \chi_{1B}\rangle= \langle\Delta^2 \chi_{1A}\rangle = \omega^2$, the non-classicality of the resource can be read off in the range $0\leq \omega <\sqrt{(1-e^{-2r})}$. It is evident that for projective measurements, $\omega\rightarrow 0$, the measurement can reveal the exact inseparability criterion.

Note that the estimation of the nonclassicality or inseparability measure $\zeta$ in Eqs.~(\ref{eq:q_fn}) and (\ref{eq:one-round}) depends on quadrature variances and, ideally, requires infinite copies of the resource state. However, in practical settings, only a finite sample, $N$, of resource states is available. The finite sampling error in measuring $\zeta$ is of the order of $\Delta\zeta \sim 1/\sqrt{N}$. A more detailed description of the sampling error is provided in Appendix~\ref{app:error}. To account for the sampling error, the detection criterion for entanglement must be $\zeta < 2 - \Delta\zeta$, which yields the condition on the minimum number of resource states required, $N \sim 1/(2-\zeta)^2$. This implies that for states with $\zeta \sim 0$ (the TMSV state with $r\rightarrow \infty$), the required resource quantity, $N$, for entanglement detection is quite low, as compared to states where $\zeta \sim 2$. If an estimate of the nonclassicality of a state, $\zeta$, is {\it a priori} available, then doing many rounds of sharp measurement using fewer $N$ in each round can be a better detection scheme. However, in typical multiparty CV communication protocols, the nonclassicality of resource state between any two modes is not known, and as such a comparable number of resource states, $N$, is required to accurately estimate $\zeta$.


\section{Sequential detection of non-classicality}
\label{sec:seq_ent}

Let us demonstrate how resource reusability using unsharp measurements provides an ideal tool to detect entanglement sequentially and independently.
Let a two-mode entangled state be initially shared between two distant parties, $A_1$ and $B_1$. Both parties independently make unsharp measurements on their quadratures using auxiliary squeezed states. The post-measurement state is then passed on to a second set of parties, $A_2$ and $B_2$. 
As shown in Fig.~\ref{fig:weak_measurement}(b), these parties then repeat the process independently and pass the post-measurement state to $(A_3, B_3)$ and so on.
The protocol aims to determine
how many pairs of parties, such as $\{(A_1, B_1), (A_2, B_2), \ldots, (A_n, B_n)\}$, can sequentially detect that their state is entangled, as determined by the inseparability criterion $\zeta(A_i, B_i)$. 
Let $\mathcal{D}_n$ be the total number of such sequential detections of entanglement in the protocol. 

When the initial shared resource is the TMSV state, the inseparability criterion after round $n$ of unsharp measurements is given as (see Appendix.~\ref{app:sequential})
\begin{eqnarray}
     \nonumber\zeta(A_n,B_n)&=&\left[\frac{1}{2}\sum_{i=1}^{n-1}(\langle\Delta^2 \chi_{2A_i}\rangle+\langle\Delta^2 \chi_{2B_i}\rangle) +2e^{-2r}\right]\\
     &+&\langle\Delta^2 \chi_{1A_n}\rangle+\langle\Delta^2 \chi_{1B_n}\rangle
     \label{eq:gen_ent}.
\end{eqnarray}
In the above equation, if no measurement is performed in the initial $n-1$ rounds, i.e., $\langle\Delta^2 \chi_{1 k_i}\rangle \to \infty \implies \langle\Delta^2 \chi_{2 k_i}\rangle \to 0 ~ \forall ~ i < n$, and a perfect homodyne measurement is employed by each party in the round $n$ with $\langle\Delta^2 \chi_{1 k_n}\rangle \to 0$, we recover the inseparability criterion for the TMSV state, $\zeta(A ,B) = 2 e^{-2r}$. We now analyze the performance of this unsharp measurement-based entanglement detection scheme with respect to two distinct situations.

\noindent \emph{(i) Equal unsharp parameter.--} Consider a scenario where each pair of parties sequentially detects the entanglement of the shared resource using identical parameters of unsharpness, i.e., $\langle\Delta^2 \chi_{1 A_i}\rangle=\langle\Delta^2 \chi_{1 B_i}\rangle=\omega_i^2 = \omega^2\;\; \forall i$.
In this situation, $\mathcal{D}_n$ turns out to be a function of $r$, and $\omega$, and can be derived from Eq.~\eqref{eq:gen_ent} as

\begin{eqnarray}
    \mathcal{D}_n< 1-8\omega^2(\omega^2-(1-e^{-2r})).
\end{eqnarray}
This provides an upper bound on $\mathcal{D}_n$. To find the maximum value of $\mathcal{D}_n$, the above equation can be differentiated with respect to $\omega^2$ which gives $\omega_{\text{opt}}=\sqrt{e^{-r}\sinh{r}}$, and  $\mathcal{D}_n^{\max}$ can be expressed as 
 \begin{eqnarray}
     \mathcal{D}_n^{\max}  < 3-2e^{-2r}(2-e^{-2r}).
 \end{eqnarray}
This clearly demonstrates that for any squeezing strength of the initial resource in the range $\frac{1}{2}\ln(2+\sqrt{2})\leq r \leq \infty$,  $\mathcal{D}_n^{\max}=2$, while for $0<r<\frac{1}{2}\ln(2+\sqrt{2})$, it is unity. 

\noindent \emph{(ii) Different unsharp parameter in each round.--} Let us explore a situation where each pair of parties has different unsharp parameters, i.e., $\omega_i^2\neq\omega_j^2$ for $i \neq j$. The goal of this scenario is to determine the maximum detection number \(\mathcal{D}_n^{\text{max}}\) given that \(\zeta(A_i, B_i) < 2\) for all \(i \leq n\). In this scenario, by applying Eq.~\eqref{eq:gen_ent}, entanglement can be detected sequentially for up to $N$ rounds if

\begin{eqnarray}
    \sum_{i=1}^{n-1}\frac{1}{4\omega_i^2}+2(\omega_n^2+e^{-2r})<2 ~~\forall n\leq N.
    \label{eq:diff_parameter}
\end{eqnarray}

Let us consider $r \to \infty$, which corresponds to the maximally entangled state. Moreover, for $r \to \infty$, the $r$-dependent term on the left-hand side of Eq. (\ref{eq:diff_parameter}) has the smallest contribution, facilitating entanglement detection for the maximum number of rounds. The necessary condition to detect entanglement sequentially up to $N$ rounds can be determined as follows. Eq.~\eqref{eq:diff_parameter} illustrates $\omega_1^2<1$ for $n=1$. By substituting the upper bound of $\omega_1^2$ in Eq.~\eqref{eq:diff_parameter} for $n=2$, we find that $\omega_2^2<7/8$. Similarly, it can be shown that  $\omega_3^2<0.732142$, $\omega_4^2<0.561411$, $\omega_5^2<0.338758$, and $\omega_6^2<-0.0302$.
Since, being the variance of a physical quantity, $\omega^2_{i}\geq0$ $\forall i$, the necessary condition dictates that we cannot detect entanglement in round $6$. 

Let us find whether there exists any explicit configuration for which entanglement can be detected up to $5$ rounds. Consider a scenario where \(\zeta(A_i, B_i) = \zeta \to 2^- ~\forall~ i\). In this configuration, the unsharpeness parameter of round $n+1$ can be rewritten as 

\begin{eqnarray}
    \omega^2_{n+1}=\frac{\zeta_{\mathcal{D}_n}-2e^{-2r}}{2}-\frac{1}{8}\sum_{j=1}^n\frac{1}{\omega^2_j}.
    \label{eq_equal_nonclassicality}
\end{eqnarray}  

\begin{figure}
\centering
\includegraphics[width=0.8\linewidth]{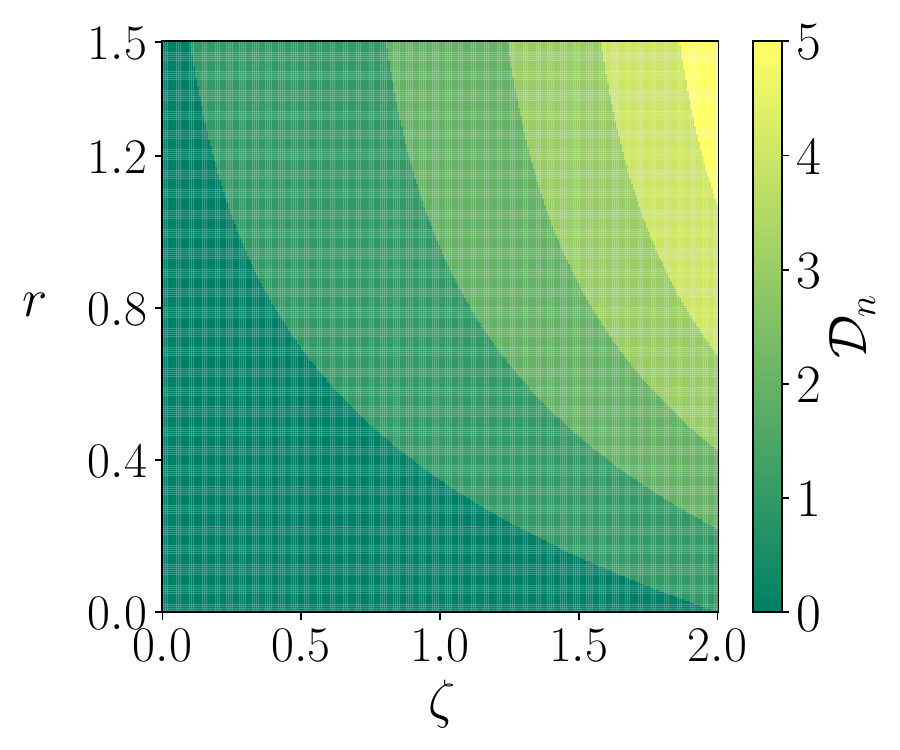}
\caption{Color map for sequential detection number, $\mathcal{D}_n$, against the squeezing strength, $r$ (ordinate), and equal entanglement, $\zeta$ (abscissa) for different unsharp parameters in every round.} 
\label{fig:equal_nonclassical} 
\end{figure}

\noindent It is clear from the above equation that there exists a range of $\zeta$ (and consequently of the unsharp parameters) which depends on the squeezing strength, $r$, such that $\mathcal{D}_n$ remains constant throughout the range as illustrated in  Fig.~\ref{fig:equal_nonclassical}. It is observed that $\mathcal{D}_n^{\text{max}}=5$ for any squeezing, $r>1.07$, given that $\zeta \rightarrow 2^-$. Note that, for any fixed $r$, there is a minimum $\zeta$ for which $\mathcal{D}_n=0$, as shown in Fig.~\ref{fig:equal_nonclassical}. This happens because we cannot detect the quantity Eq.~\eqref{eq:one-round} beyond its minimum value $2e^{-2r}$, where the minimum value is obtained from the sharp measurement.
In Table.~\ref{tab:equal_nonclassical}, we report the relation between $\mathcal{D}_n^{max}$ and squeezing $r$ at $\zeta_{\mathcal{D}_n}\rightarrow 2^-$. Note that our configuration may not be optimal; one may consider a different configuration and measurement protocol to detect entanglement for a higher number of rounds.

\begin{table}[] 
\begin{center}
\begin{tabular}{|c|c|}
\hline
     Range of $r$  & $\mathcal{D}_n$ \\\hline
      $1.0706-\infty$& 5\\\hline
      $0.6752- 1.0706$& 4\\\hline
      $0.4245-0.6751$&3\\\hline
      $0.2181-0.4244$&2\\\hline
      $0.0001-0.2181$&1\\\hline
\end{tabular}
\caption{Sequential detecting number, $\mathcal{D}_n$ with  squeezing, $r$ when $\zeta\rightarrow 2^-$ for different unsharp parameter in each round.}
\label{tab:equal_nonclassical}
\end{center}
\end{table}

We note that the parties sharing the resource at any point are oblivious to the resource squeezing in any realistic setting, thereby making them incapable of tuning the unsharp parameters accordingly. 
Nevertheless, simply from the point of view of resource reusability,
it suffices to demonstrate the power of unsharp measurements in detecting CV entanglement over multiple rounds.

\section{Conclusion}
\label{sec:conclu}
The resources required for quantum information processing protocols are generally hard to generate. From maximally entangled states for discrete variable communication to highly squeezed and non-Gaussian states for continuous variable protocols, the deterministic and accurate generation of the required states is experimentally challenging. 
On the other hand, a majority of quantum information protocols depend on projective measurements during their implementation,
which completely destroys the resource state to maximize the potential outcome.
However, an immediate problem arises if the protocol is not completed post measurement. The initial resource state is no longer usable, and more resources need to be generated for further implementation of the protocol.
One way to overcome this limitation is to create approximate clones or distill the maximal resource from multiple imperfect copies, but such schemes either yield states that are not resourceful or only a limited number of resources. In discrete variable systems, weak or unsharp measurements are a potential solution where the resource is not completely purged at the cost of obtaining a non-optimal outcome. However, the key benefit is that a single resource state can be employed multiple times in an information processing task while providing a necessary trade-off between the disturbance of the resource and the achievable quantum advantage. 

In this work, we demonstrated two schemes for the reusability of CV resources for
the sequential implementation of quantum information theoretic protocols. 
The resource-splitting scheme comprises the creation of multiple states possessing lower resource content, starting from an initial resource state. These split copies can then be utilized to sequentially carry on the protocol more than once, provided that the state possesses just enough resource to achieve quantum advantage at each round. Moreover, the resource-splitting scheme can be achieved using only free states and linear optical elements. 
The second strategy we demonstrated is the weak or unsharp quadrature measurement for CV states~\cite{Das_PRA_2014, Das_JPA_2017}.
The scheme employs the measurement of an auxiliary state instead of directly measuring the state, thereby conserving the inherent resource for further implementation of the concerned information processing task.

We exhibited the usefulness of resource reusability by using it in sequential quantum information protocols. 
For instance, the resource-splitting scheme is applied to the CV teleportation protocol, where we observed that the maximum number of sequential teleportation attempts that can yield non-classical fidelity is dependent on the parameters of the state and the resource-splitting mechanism. 
We investigated the performance of the protocol for different rounds by considering two constraints - $(1)$ when equal non-classical fidelity is demanded at each round, and $(2)$ if the resource is identically split in each round. In the former case, we obtained a relationship between the minimum fidelity and the maximum number of rounds, illustrating that a significantly high number of trials can be carried out if the requisite fidelity is just above the classical threshold. On the other hand, in the latter case, we showed a trade-off between the resource-splitting parameter and the initial resource squeezing while obtaining the maximum number of rounds with quantum advantage. We showed that a lower initial resource content required more stringent bounds on the resource-splitting parameter, and also, the maximum round number in this scenario is significantly lower than the case with equal fidelity. 

In CV quantum communication protocols, resources with non-classicality, based on relative quadrature variances, have been shown to demonstrate quantum advantage~\cite{Das_PRA_2024}. Therefore, for any unknown multimode state, it is important to determine the non-classicality without significantly disturbing the system.
We showed that our resource reusability approach based on the unsharp measurements of quadratures is an important tool for detecting the entanglement of a Gaussian resource sequentially.
We observed that there is a limit on the number of parties that can sequentially detect
entanglement
using unsharp measurements on a given resource, similar to unsharp measurements in discrete variable entanglement detection protocols. By using an equal unsharp parameter at every round, entanglement can be sequentially detected twice for a finite value of resource squeezing. 
Similarly, a different unsharp parameter at each round can lead to a maximum detection number of five, provided the resource squeezing strength is greater than $1.07$. This helped us to put forward the remarkable ability of unsharp measurements to allow the reiteration of protocols using only single-mode auxiliary states and Gaussian measurements.

Although studies on resource reusability and sequential quantum protocols have been extensively conducted in discrete variable systems, the investigation of such properties in continuous variable systems is fairly limited.
However, in recent years, CV platforms have emerged as one of the leading candidates for realizing several quantum communication, cryptographic, and computation protocols. As such, our work addresses an important question related to the implementation of sequential quantum protocols by reusing expensive resource states, which can be adapted to several CV communication protocols.
Our work paves the way for further research concerning the application of resource-splitting procedures in other aspects of CV quantum information theory and photonic platforms for quantum computation. 

\section*{Acknowledgement} 
This research was supported in part by the ``INFOSYS scholarship for senior students''. R.G. acknowledges funding from the HORIZON-EIC-$2022$-PATHFINDERCHALLENGES-$01$ program under Grant Agreement No.~$10111489$ (Veriqub). Views and opinions expressed are, however, those of the authors only and do not necessarily reflect those of the European Union. Neither the European Union nor the granting authority can be held responsible for them. H.S.D. acknowledges support from SERB-DST, India, under a Core-Research Grant (No: CRG$/2021/008918$) and funding
from IRCC, IIT Bombay (No: RD$/0521-$ IRCCSH$0-001$). 

\appendix
\section{CV teleportation under resource-splitting scheme}
\label{app:recycle_protocol}
Consider that a TMSV state, $\rho_{BA}$, is initially shared between two parties, Alice ($A$) and Bob ($B$). The covariance matrix (CM), $\sigma_{BA}$, and the displacement vector of first-order moments, $\mathbf{r}_{BA}$, of the state are respectively given by 

\begin{eqnarray}
    \sigma_{BA}=\begin{pmatrix}
         \cosh{2r}&0&\sinh{2r}&0\\
        0&\cosh{2r}&0&-\sinh{2r}\\
        \sinh{2r}&0&\cosh{2r}&0\\
        0&-\sinh{2r}&0&\cosh{2r}
    \end{pmatrix},
    \label{eq_tmsv_cm}
\end{eqnarray}    
\begin{eqnarray}
    \mathbf{r}_{BA}=\begin{pmatrix}
        0\\
        0\\
        0\\
        0
    \end{pmatrix}.
\end{eqnarray}
For Gaussian states, the resource splitting protocol mentioned in Sec.~\ref{sec:re-splitting} can be represented in terms of the CM and the displacement vector. In this case, the unitary evolution of the resource state with auxiliary vacuum modes corresponds to the symplectic transformation of the corresponding CMs and the displacement vectors. 

To reuse the resource, Alice combines her mode with a vacuum mode having CM, $\sigma_{v_A}=\mathbb{I}_2$ ($\mathbb{I}_2$ being the two-dimensional identity matrix), and displacement vector, $\mathbf{r}_{v_A}=(0,0)^T$. Therefore, Eq.~\eqref{eq_splitting_res} turns out to be 

\begin{eqnarray}
    \nonumber\sigma_{BA_TA_R}&=& \mu(\tau_A) \sigma_{BAv_A} \mu^T(\tau_A),\\
    r_{BA_TA_R}&=&\mu(\tau_A) \mathbf{r}_{BAv_A},
    \label{eq:state_transform_cov}
\end{eqnarray}
where $\sigma_{BAv_A}=\sigma_{BA} \bigoplus \sigma_{v_A}$, and $\mathbf{r}_{BAv_A}= \mathbf{r}_{BA} \bigoplus \mathbf{r}_{v_A}$, while $\mu(\tau_A)=\mathbb{1}_2 \bigoplus BS(\tau_A)$, with $BS(\tau_A)$ being the matrix representation of a beam splitter of transmitivity $\tau_A$, given as
\begin{eqnarray}
    \nonumber BS(\tau_A)=\begin{pmatrix}
         \sqrt{\tau_A}&0&\sqrt{1-\tau_A}&0\\
        0&\sqrt{\tau_A}&0&\sqrt{1-\tau_A}\\
        \sqrt{1-\tau_A}&0&-\sqrt{\tau_A}&0\\
        0&\sqrt{1-\tau_A}&0&-\sqrt{\tau_A}
    \end{pmatrix}.\\
\end{eqnarray}
Therefore, the CM and the displacement vector of the transmitted (reflected) resource, $\rho_{BA_T} (\rho_{BA_R})$, can be written as $\sigma_{BA_T} (\sigma_{BA_R})$, and $\mathbf{r}_{BA_T} (\mathbf{r}_{BA_R})$ by tracing out the reflected (transmitted) mode. Using the transmitted twin-beam $\rho_{BA_T} \equiv \{\mathbf{r}_{BA_T}, \sigma_{BA_T}\}$, suppose Alice wants to teleport a coherent state, $\rho_{\alpha}$, having covariance matrix, $\sigma_{\alpha}$, and displacement vector, $\mathbf{r}_{\alpha}$, given by 

\begin{eqnarray}
    \sigma_{\alpha}=\begin{pmatrix}
        1&0\\
        0&1
    \end{pmatrix},
    \mathbf{r}_{\alpha}=\begin{pmatrix}
        x_\alpha\\
        p_\alpha
    \end{pmatrix}.
\end{eqnarray}
Thus, the total CM and displacement vector of the initial composite state, $\rho=\rho_{BA}\otimes\rho_\alpha$ is $\{\sigma=\sigma_{BA_T}\bigoplus\sigma_{\alpha}, \mathbf{r} = \mathbf{r}_{BA_T} \bigoplus \mathbf{r}_{\alpha}\}$.  Alice will now perform a double homodyne measurement on her two modes, $A_T$ and $\alpha$. Due to the projective measurement, the conditional state at Bob's end can be given as~\cite{Serafini_2017}  

\begin{eqnarray}
    \nonumber\sigma_B &\longrightarrow& \sigma_B- \sigma_{B:(A_T \alpha)} (\sigma_{A_T\alpha}+\sigma_P)^{-1} \sigma^T_{B:(A_T\alpha)}\\
    \mathbf{r}_B &\longrightarrow& \mathbf{r}_B + \sigma_{B:(A_T\alpha)}.(\sigma_{A_T\alpha}+\sigma_P)^{-1} (\mathbf{r}_P-\mathbf{r}_B).~~~~
    \label{eq:displacement}
\end{eqnarray}
In the above equation, $\sigma_{B:(A_T\alpha)}$ is the correlation matrix between two subsystems $\rho_B$ and $\rho_{A_T\alpha}$, $\sigma_{A_T \alpha}$ is the covariance matrix of the two modes at Alice's station,  and
 $\{\mathbf{r}_P,\sigma_P\}$ is the displacement vector and CM of an Einstein-Podolski-Rosen (EPR) state~\cite{Braunstein_RMP_2005}, expressed as

\begin{eqnarray}
    \sigma_P=\lim_{cc\rightarrow\infty}\begin{pmatrix}
        cc\;\mathbb{1}_2 & cc\;\sigma_z\\
        cc\;\sigma_z & cc\;\mathbb{1}_2
    \end{pmatrix}, \mathbf{r}_P=\begin{pmatrix}
        0\\
        \mathbf{r}_M
    \end{pmatrix},
    \label{eq:bell-measurement}
\end{eqnarray}
with $\mathbf{r}_M$ being the displacement vector of measurement outcomes. Taking the limit $cc\rightarrow \infty$ in Eq.~\eqref{eq:displacement} corresponds to performing a Bell measurement or a double-homodyne measurement. The measurement, $\mathbf{r}_M$, occurs with the probability 

\begin{eqnarray}
    P(\mathbf{r}_M)=\frac{e^{-(\mathbf{r}-\mathbf{r}_{A_T \alpha})^T(\sigma_{A_T \alpha}+\sigma_P)^{-1}(\mathbf{r}_M-\mathbf{r}_{A_T \alpha})}}{\pi\sqrt{\text{Det}(\sigma_{A_T \alpha}+\sigma_P)}}.
    \label{eq:prob}
\end{eqnarray}
Displacement of an amount $\mathbf{r}_M$ in $\mathbf{r}_B$, i.e., $\mathbf{r}_B\rightarrow \mathbf{r}_B+\mathbf{r}_M$, leads to the teleported state at the receiver's end with fidelity

\begin{eqnarray}
    \nonumber\mathcal{F}=\Tr[\rho_B\rho_\alpha]&=&\int d\mathbf{r}_M P(\mathbf{r}_M)\frac{2}{\sqrt{Det(\sigma_\alpha+\sigma_B)}}\\
    &\times&e^{-(\mathbf{r}_B-\mathbf{r}_M)^T(\sigma_M+\sigma_B)^{-1}(\mathbf{r}_B-\mathbf{r}_M)},
    \label{eq:genfid}
\end{eqnarray}
where the average is taken over all possible outcomes of the double-homodyne measurement. The explicit form of the fidelity is given in Eq.~\eqref{eq:first_round}.
If Bob is unable to complete the protocol, Alice can reuse the reflected beam as a new shared resource, i.e., $\rho_{BA_2}=\rho_{BA_R}$. Performing all the steps again from Eq.\eqref{eq:state_transform_cov}, the fidelity, $\mathcal{F}_2$ of the second round can be found. The fidelity of the round $n$ is described in Eq.\eqref{eq:genfid_0}.

\section{Sequential detection of entanglement using unsharp measurements}
\label{app:sequential}
Our aim is to determine the non-classicality measure $\zeta(A_1,B_1)$ of the shared state, mentioned in Eq.~\eqref{eq:one-round}, using the unsharp homodyne measurement protocol as described in Sec.~\ref{sec:weak_meas}. 

Let us consider two parties, $A_1$ and $B_1$, who share a TMSV state. The covariance matrix (CM) of the TMSV state can be expressed as $\sigma_{A_1 B_1}$ (see Eq.~\eqref{eq_tmsv_cm}) with quadratures $\eta_{ik}=\{\hat{q}_{k}, \hat{p}_{k}\}$ where the index $k$ denotes the parties, i.e., $k=\{A_1, B_1\}$ and $i = \{1, 2\} \equiv \{\hat{q}, \hat{p}\}$. Then, the Wigner function of the shared resource is given as 

\begin{eqnarray}
    \mathcal{W}(\eta_{i A_1},\eta_{j B_1})=\frac{1}{\pi\sqrt{\det(\sigma_{A_1B_1})}}e^{-\eta^T [\sigma_{A_1B_1}]^{-1} \eta}.
\end{eqnarray}
In the above equation, $\eta = \{\eta_{i A_1}, \eta_{j B_1}\}$. To measure the quadratures of the shared resource, both parties combine their respective mode with auxiliary states  $\rho_{aA_1}$ and $\rho_{aB_1}$, respectively. The CM of each auxiliary state is given by

\begin{eqnarray}
    \sigma_{ak}=\begin{pmatrix}
        \langle\Delta^2 \chi_{1k}\rangle &0\\
        0& \langle\Delta^2 \chi_{2k}\rangle
    \end{pmatrix},
\end{eqnarray}
with $\chi_{ik}=\{q_{ak}, p_{ak}\}$. The Wigner function of the composite system can thus be written as 

\begin{eqnarray}
    \mathcal{W}(\nu_{A_1},\nu_{B_1})=\mathcal{W}(\chi_{i A_1})\mathcal{W}(\eta_{j A_1},\eta_{l B_1})\mathcal{W}(\chi_{s B_1}),
    \label{eq_initial_wigner_fn}
\end{eqnarray}
where $\mathcal{W}(\chi_{i k})$ is the Wigner function of the auxiliary system $\rho_{ak}$, and $\nu_{k}=\{\eta_{i k},\chi_{j k}\}$ is the joint quadrature vector of the mode, $k$, and its corresponding auxiliary subsystem. In the Gaussian regime, the unitary evolution of a Gaussian state corresponds to a symplectic transformation on both its covariance matrix and displacement vector. The symplectic matrix corresponding to the Hamiltonian mentioned in Eq.~\eqref{eq_hamiltonian} is $\mathcal{S}_{\eta_{ik}}=e^{\Omega H_{\eta_{ik}}} $ with $\mathcal{H}_{\eta_{ik}}=\frac{1}{2} \nu^T_{k} H_{\eta_{ik}} \nu_{k}$~\cite{Serafini_2017}. The two-mode symplectic form $\Omega$ is given by 
\begin{eqnarray}
    \Omega=\begin{pmatrix}
        0&1&0&0\\
        -1&0&0&0\\
        0&0&0&1\\
        0&0&-1&0
    \end{pmatrix},
\end{eqnarray}
and explicit form of the symplectic transformations, $\mathcal{S}$, are 
\begin{eqnarray}
    \mathcal{S}_{\eta_{1k}}=\begin{pmatrix}
        1&0&0&0\\
        0&1&0&-1\\
        1&0&1&0\\
        0&0&0&1
    \end{pmatrix};~~ \mathcal{S}_{\eta_{2k}}=\begin{pmatrix}
        1&0&0&1\\
        0&1&0&0\\
        0&1&1&0\\
        0&0&0&1
    \end{pmatrix}.
\end{eqnarray}
If the quadratures measured are $\eta_{iA}$ and $\eta_{jB}$, under the corresponding symplectic transformations, the quadrature for each subsystem, i.e., $\nu_{k}$ changes as  
 \(  \nu_{k}'=\mathcal{S}_{\eta_{ik}}\nu_{k}\),
and the corresponding total Wigner function, Eq.~\eqref{eq_initial_wigner_fn}, for this particular measurement setting, turns out to be
\begin{eqnarray}
    \nonumber\mathcal{W}(\nu_{A_1},\nu_{B_1})&&\xrightarrow{\mathcal{S}_{\eta_{iA_1}}\bigoplus\mathcal{S}_{\eta_{jB_1}}}\mathcal{W}'(\nu_{A_1}',\nu_{B_1}')_{(\eta_{iA_1},\eta_{jB_1})}\\
    &\equiv&\mathcal{W}(\mathcal{S}_{\eta_{iA_1}}^{-1}\nu_{A_1}',\mathcal{S}_{\eta_{jB_1}}^{-1}\nu_{B_1}')_{(\eta_{iA_1},\eta_{jB_1})}.
    \label{eq:change_wig}
\end{eqnarray}
The unnormalized conditional Wigner function of the post-measurement state can be determined by integrating over the two auxiliary subsystem quadratures
\begin{widetext}
\begin{eqnarray}
\mathcal{W}'(\eta_{lA_1}',\eta_{sB_1}')_{(\eta_{iA_1},\eta_{jB_1})}=\int\mathcal{W}'(\nu_{A_1}',\nu_{B_1}')_{(\eta_{iA_1},\eta_{jB_1})}\times\delta(\chi_{1A_1}'-\chi_{1A_1}'^{m})\times\delta(\chi_{1B_1}'-\chi_{1B_1}'^{m})\prod_{i,k}d\chi_{ik}',
\label{eq_post_measurement}
\end{eqnarray}
\end{widetext}
where $\chi'^m_{1A_1}, \chi'^m_{1B_1}$ are the measurement outcomes corresponding to the quadratures  $(\chi_{1A_1}',\chi_{1B_1}')$. Therefore, the Wigner function of  unconditional normalized post-measurement state can be written as
\begin{eqnarray}
\nonumber&&\mathcal{W}'(\eta_{lA_1}',\eta_{sB_1}')_{(\eta_{iA_1},\eta_{jB_1})}\\
&=&\int\mathcal{W}'(\nu_{A_1}',\nu_{B_1}')_{(\eta_{iA_1},\eta_{jB_1})}d\chi_{ik}'.
\label{eq_post_measurement}
\end{eqnarray}
The above Wigner function can be found by averaging over all possible measurement outcomes $\chi'^m_{1A_1}, \chi'^m_{1B_1}$.The corresponding probability of the measurement outcome is given as

\begin{eqnarray}
    \nonumber && P(\chi_{1A}',\chi_{1B}')_{(\eta_{iA_1},\eta_{jB_1})}\\
    &=&\int \mathcal{W}'(\nu_{A_1}',\nu_{B_1}')_{(\eta_{iA_1},\eta_{jB_1})}\prod_{ k}d\chi_{2 k}'\prod_{j, k'} d\eta_{j k'}'.
\end{eqnarray}
To determine the inseparability criterion of the shared resource, $\zeta(A_1, B_1)$, one only needs to calculate the the probability function, $P(\chi_{1A}',\chi_{1B}')_{\eta_{1A_1},\eta_{1B_1}}$ as demonstrated in Eqs.~\eqref{eq_vanloock_cal} and~\eqref{eq:van_loock_prob}. On the other hand, both parties, $A_1$ and $B_1$, can independently measure any of their quadratures, i.e., $(\eta_{iA_1}',\eta_{jB_1}')$ where $i,j=\{1,2\}$ with equal probability. Thus, the average post-measurement state is the ensemble of four states, and its corresponding Wigner function can be expressed as a convex combination of Eq.~\eqref{eq_post_measurement} for each realization of $(i,j)$, such that
\begin{eqnarray} 
\omega(\eta_{l A_2},\eta_{s B_2})
&=&\frac{1}{4}\sum_{i,j}\mathcal{W}'(\eta_{l A_1}',\eta_{s B_1}')_{(\eta_{iA_1},\eta_{jB_1})}.
\label{eq_post_measurement_state}
\end{eqnarray}
In the above equation, $(\eta_{l A_2},\eta_{s B_2})$ represents the quadratures of the average post-measurement state belonging to the second pair of parties, $(A_2,B_2)$. The two parties can independently repeat the process of detecting the entanglement of their post-measurement shared resource. In this case, inserting the Wigner function, mentioned in  Eq.\eqref{eq_post_measurement_state} into Eq.~\eqref{eq_initial_wigner_fn} and calculating the corresponding probabilities, the inseparability criterion, $\zeta(A_2,B_2)$, can be evaluated. Further rounds of calculations proceed recursively to yield Eq.~\eqref{eq:gen_ent}.

\section{Statistical error in calculating quadrature variance}
\label{app:error}

The standard deviation or variance of an optical field quadrature    $\hat{\mathcal{O}}$ is derived using the relation~\cite{Serafini_2017}, 
%
\begin{eqnarray}
\langle\Delta^2\hat{\mathcal{O}}\rangle &=&\langle\hat{\mathcal{O}^2}\rangle-\langle\hat{\mathcal{O}}\rangle^2 \nonumber\\
    &=&\int \mathcal{O}^2P(\mathcal{O})d\mathcal{O}-\left(\int \mathcal{O}P(\mathcal{O})d\mathcal{O}\right)^2,~
    \label{theory}
\end{eqnarray}
where, $P(\mathcal{O})$ is the probability distribution of measurement outcomes in the phase space. In principle, to estimate the variance $\langle\Delta^2\hat{\mathcal{O}}\rangle$, one needs to perform an infinitely large number of homodyne measurements. However, in a practical setting the optical field is sampled only a finite $N$ number of times, such that
\begin{eqnarray}
&\langle\Delta^2\hat{\mathcal{O}}\rangle_{\text{finite}} =\sum_i^N\mathcal{O}_i^2P(\mathcal{O}_i)-\left(\sum_i^N\mathcal{O}_iP(\mathcal{O}_i)\right)^2,& \nonumber\\
&\lim_{N\to\infty}\langle\Delta^2\hat{\mathcal{O}}\rangle_{\text{finite}}=\langle\Delta^2\hat{\mathcal{O}}\rangle.&
    \label{measured}
\end{eqnarray}
%
%
The quantity $\langle\Delta^2\hat{\mathcal{O}}\rangle_{\text{finite}}$ 
converges to its true value 
for a large $N$.
%

In homodyne measurement, the quadrature $\mathcal{O}_{\text{meas}}$ is obtained in every measurement.  Let us consider that the error of the quadrature from its actual mean value is given by $\sigma$ when the measurement is repeated $N$ times. If the actual distribution of the quadrature $\mathcal{O}$ is Gaussian with mean $\langle\mathcal{O}\rangle$ and variance $\langle\Delta^2\hat{\mathcal{O}}\rangle$, its measured value  $\mathcal{O}_{\text{meas}}$ after $N$ samplings is also a Gaussian with variance $\sigma^2=\langle\Delta^2\hat{\mathcal{O}}\rangle/N$. A detailed derivation of the above estimation error is given in Chapter~$31$ of Ref.~\cite{Riley_Hobson_Bence_2006}.
Therefore, the estimation error is given by the relation
\begin{eqnarray}
    \sigma \sim\frac{1}{\sqrt{N}}.
\end{eqnarray}

For our nonclassicality measure $\zeta(A,B)$, we need to calculate the variance of the quadrature difference between two modes, given by $\hat{\mathcal{O}}=\chi_{1A}-\chi_{1B}$ (and, $\chi_{1A}+\chi_{1B}$) using $P(\hat{\mathcal{O}})$, which is the probability distribution of the measurement outcomes $(\chi_{1A},\chi_{1B})$ as stated in Eq.~\eqref{eq:van_loock_prob}.

\bibliography{ref.bib}

\begin{thebibliography}{99}%
\makeatletter
\providecommand \@ifxundefined [1]{%
 \@ifx{#1\undefined}
}%
\providecommand \@ifnum [1]{%
 \ifnum #1\expandafter \@firstoftwo
 \else \expandafter \@secondoftwo
 \fi
}%
\providecommand \@ifx [1]{%
 \ifx #1\expandafter \@firstoftwo
 \else \expandafter \@secondoftwo
 \fi
}%
\providecommand \natexlab [1]{#1}%
\providecommand \enquote  [1]{``#1''}%
\providecommand \bibnamefont  [1]{#1}%
\providecommand \bibfnamefont [1]{#1}%
\providecommand \citenamefont [1]{#1}%
\providecommand \href@noop [0]{\@secondoftwo}%
\providecommand \href [0]{\begingroup \@sanitize@url \@href}%
\providecommand \@href[1]{\@@startlink{#1}\@@href}%
\providecommand \@@href[1]{\endgroup#1\@@endlink}%
\providecommand \@sanitize@url [0]{\catcode `\\12\catcode `\$12\catcode `\&12\catcode `\#12\catcode `\^12\catcode `\_12\catcode `\%12\relax}%
\providecommand \@@startlink[1]{}%
\providecommand \@@endlink[0]{}%
\providecommand \url  [0]{\begingroup\@sanitize@url \@url }%
\providecommand \@url [1]{\endgroup\@href {#1}{\urlprefix }}%
\providecommand \urlprefix  [0]{URL }%
\providecommand \Eprint [0]{\href }%
\providecommand \doibase [0]{http://dx.doi.org/}%
\providecommand \selectlanguage [0]{\@gobble}%
\providecommand \bibinfo  [0]{\@secondoftwo}%
\providecommand \bibfield  [0]{\@secondoftwo}%
\providecommand \translation [1]{[#1]}%
\providecommand \BibitemOpen [0]{}%
\providecommand \bibitemStop [0]{}%
\providecommand \bibitemNoStop [0]{.\EOS\space}%
\providecommand \EOS [0]{\spacefactor3000\relax}%
\providecommand \BibitemShut  [1]{\csname bibitem#1\endcsname}%
\let\auto@bib@innerbib\@empty
\bibitem [{\citenamefont {Horodecki}\ \emph {et~al.}(2009)\citenamefont {Horodecki}, \citenamefont {Horodecki}, \citenamefont {Horodecki},\ and\ \citenamefont {Horodecki}}]{Horodecki_RMP_2009}%
  \BibitemOpen
  \bibfield  {author} {\bibinfo {author} {\bibfnamefont {R.}~\bibnamefont {Horodecki}}, \bibinfo {author} {\bibfnamefont {M.}~\bibnamefont {Horodecki}}, \bibinfo {author} {\bibfnamefont {K.}~\bibnamefont {Horodecki}}, \ and\ \bibinfo {author} {\bibfnamefont {P.}~\bibnamefont {Horodecki}},\ }\href {\doibase 10.1103/RevModPhys.81.865} {\bibfield  {journal} {\bibinfo  {journal} {Rev. Mod. Phys.}\ }\textbf {\bibinfo {volume} {81}},\ \bibinfo {pages} {865} (\bibinfo {year} {2009})}\BibitemShut {NoStop}%
\bibitem [{\citenamefont {Das}\ \emph {et~al.}(2016)\citenamefont {Das}, \citenamefont {Chanda}, \citenamefont {Lewenstein}, \citenamefont {Sanpera}, \citenamefont {De},\ and\ \citenamefont {Sen}}]{Das_QI_2016}%
  \BibitemOpen
  \bibfield  {author} {\bibinfo {author} {\bibfnamefont {S.}~\bibnamefont {Das}}, \bibinfo {author} {\bibfnamefont {T.}~\bibnamefont {Chanda}}, \bibinfo {author} {\bibfnamefont {M.}~\bibnamefont {Lewenstein}}, \bibinfo {author} {\bibfnamefont {A.}~\bibnamefont {Sanpera}}, \bibinfo {author} {\bibfnamefont {A.~S.}\ \bibnamefont {De}}, \ and\ \bibinfo {author} {\bibfnamefont {U.}~\bibnamefont {Sen}},\ }\enquote {\bibinfo {title} {The separability versus entanglement problem},}\ \ (\bibinfo  {publisher} {Wiley},\ \bibinfo {year} {2016})\ pp.\ \bibinfo {pages} {127--174}\BibitemShut {NoStop}%
\bibitem [{\citenamefont {Bennett}\ and\ \citenamefont {Wiesner}(1992)}]{Bennett_PRL_1992}%
  \BibitemOpen
  \bibfield  {author} {\bibinfo {author} {\bibfnamefont {C.~H.}\ \bibnamefont {Bennett}}\ and\ \bibinfo {author} {\bibfnamefont {S.~J.}\ \bibnamefont {Wiesner}},\ }\href {\doibase 10.1103/PhysRevLett.69.2881} {\bibfield  {journal} {\bibinfo  {journal} {Phys. Rev. Lett.}\ }\textbf {\bibinfo {volume} {69}},\ \bibinfo {pages} {2881} (\bibinfo {year} {1992})}\BibitemShut {NoStop}%
\bibitem [{\citenamefont {Bennett}\ \emph {et~al.}(1993)\citenamefont {Bennett}, \citenamefont {Brassard}, \citenamefont {Crépeau}, \citenamefont {Jozsa}, \citenamefont {Peres},\ and\ \citenamefont {Wootters}}]{Bennett_PRL_1993}%
  \BibitemOpen
  \bibfield  {author} {\bibinfo {author} {\bibfnamefont {C.~H.}\ \bibnamefont {Bennett}}, \bibinfo {author} {\bibfnamefont {G.}~\bibnamefont {Brassard}}, \bibinfo {author} {\bibfnamefont {C.}~\bibnamefont {Crépeau}}, \bibinfo {author} {\bibfnamefont {R.}~\bibnamefont {Jozsa}}, \bibinfo {author} {\bibfnamefont {A.}~\bibnamefont {Peres}}, \ and\ \bibinfo {author} {\bibfnamefont {W.~K.}\ \bibnamefont {Wootters}},\ }\href {\doibase 10.1103/PhysRevLett.70.1895} {\bibfield  {journal} {\bibinfo  {journal} {Physical Review Letters}\ }\textbf {\bibinfo {volume} {70}},\ \bibinfo {pages} {1895} (\bibinfo {year} {1993})}\BibitemShut {NoStop}%
\bibitem [{\citenamefont {Ekert}(1991)}]{Ekert91}%
  \BibitemOpen
  \bibfield  {author} {\bibinfo {author} {\bibfnamefont {A.~K.}\ \bibnamefont {Ekert}},\ }\href {\doibase 10.1103/physrevlett.67.661} {\bibfield  {journal} {\bibinfo  {journal} {Phys. Rev. Lett.}\ }\textbf {\bibinfo {volume} {67}},\ \bibinfo {pages} {661} (\bibinfo {year} {1991})}\BibitemShut {NoStop}%
\bibitem [{\citenamefont {Gisin}\ \emph {et~al.}(2002)\citenamefont {Gisin}, \citenamefont {Ribordy}, \citenamefont {Tittel},\ and\ \citenamefont {Zbinden}}]{Gisin_RMP_2002}%
  \BibitemOpen
  \bibfield  {author} {\bibinfo {author} {\bibfnamefont {N.}~\bibnamefont {Gisin}}, \bibinfo {author} {\bibfnamefont {G.}~\bibnamefont {Ribordy}}, \bibinfo {author} {\bibfnamefont {W.}~\bibnamefont {Tittel}}, \ and\ \bibinfo {author} {\bibfnamefont {H.}~\bibnamefont {Zbinden}},\ }\href {\doibase 10.1103/RevModPhys.74.145} {\bibfield  {journal} {\bibinfo  {journal} {Rev. Mod. Phys.}\ }\textbf {\bibinfo {volume} {74}},\ \bibinfo {pages} {145} (\bibinfo {year} {2002})}\BibitemShut {NoStop}%
\bibitem [{\citenamefont {Briegel}\ \emph {et~al.}(2009)\citenamefont {Briegel}, \citenamefont {Browne}, \citenamefont {Dür}, \citenamefont {Raussendorf},\ and\ \citenamefont {den Nest}}]{Briegel_Nature_2009}%
  \BibitemOpen
  \bibfield  {author} {\bibinfo {author} {\bibfnamefont {H.~J.}\ \bibnamefont {Briegel}}, \bibinfo {author} {\bibfnamefont {D.~E.}\ \bibnamefont {Browne}}, \bibinfo {author} {\bibfnamefont {W.}~\bibnamefont {Dür}}, \bibinfo {author} {\bibfnamefont {R.}~\bibnamefont {Raussendorf}}, \ and\ \bibinfo {author} {\bibfnamefont {M.~V.}\ \bibnamefont {den Nest}},\ }\href {\doibase 10.1038/nphys1157} {\bibfield  {journal} {\bibinfo  {journal} {Nature Physics}\ }\textbf {\bibinfo {volume} {5}},\ \bibinfo {pages} {19} (\bibinfo {year} {2009})}\BibitemShut {NoStop}%
\bibitem [{\citenamefont {Streltsov}\ \emph {et~al.}(2017)\citenamefont {Streltsov}, \citenamefont {Adesso},\ and\ \citenamefont {Plenio}}]{Streltsov_RMP_2017}%
  \BibitemOpen
  \bibfield  {author} {\bibinfo {author} {\bibfnamefont {A.}~\bibnamefont {Streltsov}}, \bibinfo {author} {\bibfnamefont {G.}~\bibnamefont {Adesso}}, \ and\ \bibinfo {author} {\bibfnamefont {M.~B.}\ \bibnamefont {Plenio}},\ }\href {\doibase 10.1103/RevModPhys.89.041003} {\bibfield  {journal} {\bibinfo  {journal} {Rev. Mod. Phys.}\ }\textbf {\bibinfo {volume} {89}},\ \bibinfo {pages} {041003} (\bibinfo {year} {2017})}\BibitemShut {NoStop}%
\bibitem [{\citenamefont {Deutsch}\ and\ \citenamefont {Jozsa}(1992)}]{Deutsch_PRSL_1992}%
  \BibitemOpen
  \bibfield  {author} {\bibinfo {author} {\bibfnamefont {D.}~\bibnamefont {Deutsch}}\ and\ \bibinfo {author} {\bibfnamefont {R.}~\bibnamefont {Jozsa}},\ }\href {\doibase 10.1098/rspa.1992.0167} {\bibfield  {journal} {\bibinfo  {journal} {Proceedings of the Royal Society of London. Series A: Mathematical and Physical Sciences}\ }\textbf {\bibinfo {volume} {439}},\ \bibinfo {pages} {553} (\bibinfo {year} {1992})}\BibitemShut {NoStop}%
\bibitem [{\citenamefont {Shor}(1994)}]{Shor_IEEE_1994}%
  \BibitemOpen
  \bibfield  {author} {\bibinfo {author} {\bibfnamefont {P.}~\bibnamefont {Shor}},\ }in\ \href {\doibase 10.1109/SFCS.1994.365700} {\emph {\bibinfo {booktitle} {Proceedings 35th Annual Symposium on Foundations of Computer Science}}}\ (\bibinfo  {publisher} {IEEE Computer Society Press, Washington, DC},\ \bibinfo {year} {1994})\ pp.\ \bibinfo {pages} {124--134}\BibitemShut {NoStop}%
\bibitem [{\citenamefont {Grover}(1996)}]{Grover_ACM_1996}%
  \BibitemOpen
  \bibfield  {author} {\bibinfo {author} {\bibfnamefont {L.~K.}\ \bibnamefont {Grover}},\ }in\ \href {\doibase 10.1145/237814.237866} {\emph {\bibinfo {booktitle} {Proceedings of the twenty-eighth annual ACM symposium on Theory of computing - STOC '96}}}\ (\bibinfo  {publisher} {ACM Press},\ \bibinfo {year} {1996})\ pp.\ \bibinfo {pages} {212--219}\BibitemShut {NoStop}%
\bibitem [{\citenamefont {Cleve}\ \emph {et~al.}(1998)\citenamefont {Cleve}, \citenamefont {Ekert}, \citenamefont {Macchiavello},\ and\ \citenamefont {Mosca}}]{Cleve_PRSL_1998}%
  \BibitemOpen
  \bibfield  {author} {\bibinfo {author} {\bibfnamefont {R.}~\bibnamefont {Cleve}}, \bibinfo {author} {\bibfnamefont {A.}~\bibnamefont {Ekert}}, \bibinfo {author} {\bibfnamefont {C.}~\bibnamefont {Macchiavello}}, \ and\ \bibinfo {author} {\bibfnamefont {M.}~\bibnamefont {Mosca}},\ }\href {\doibase 10.1098/rspa.1998.0164} {\bibfield  {journal} {\bibinfo  {journal} {Proceedings of the Royal Society of London. Series A: Mathematical, Physical and Engineering Sciences}\ }\textbf {\bibinfo {volume} {454}},\ \bibinfo {pages} {339} (\bibinfo {year} {1998})}\BibitemShut {NoStop}%
\bibitem [{\citenamefont {Loss}\ and\ \citenamefont {DiVincenzo}(1998)}]{Loss_PRA_1998}%
  \BibitemOpen
  \bibfield  {author} {\bibinfo {author} {\bibfnamefont {D.}~\bibnamefont {Loss}}\ and\ \bibinfo {author} {\bibfnamefont {D.~P.}\ \bibnamefont {DiVincenzo}},\ }\href {\doibase 10.1103/PhysRevA.57.120} {\bibfield  {journal} {\bibinfo  {journal} {Phys. Rev. A}\ }\textbf {\bibinfo {volume} {57}},\ \bibinfo {pages} {120} (\bibinfo {year} {1998})}\BibitemShut {NoStop}%
\bibitem [{\citenamefont {Kane}(1998)}]{Kane_Nature_1998}%
  \BibitemOpen
  \bibfield  {author} {\bibinfo {author} {\bibfnamefont {B.~E.}\ \bibnamefont {Kane}},\ }\href {\doibase 10.1038/30156} {\bibfield  {journal} {\bibinfo  {journal} {Nature}\ }\textbf {\bibinfo {volume} {393}},\ \bibinfo {pages} {133} (\bibinfo {year} {1998})}\BibitemShut {NoStop}%
\bibitem [{\citenamefont {Mills}\ \emph {et~al.}(2022)\citenamefont {Mills}, \citenamefont {Guinn}, \citenamefont {Gullans}, \citenamefont {Sigillito}, \citenamefont {Feldman}, \citenamefont {Nielsen},\ and\ \citenamefont {Petta}}]{Mills_SA_2022}%
  \BibitemOpen
  \bibfield  {author} {\bibinfo {author} {\bibfnamefont {A.~R.}\ \bibnamefont {Mills}}, \bibinfo {author} {\bibfnamefont {C.~R.}\ \bibnamefont {Guinn}}, \bibinfo {author} {\bibfnamefont {M.~J.}\ \bibnamefont {Gullans}}, \bibinfo {author} {\bibfnamefont {A.~J.}\ \bibnamefont {Sigillito}}, \bibinfo {author} {\bibfnamefont {M.~M.}\ \bibnamefont {Feldman}}, \bibinfo {author} {\bibfnamefont {E.}~\bibnamefont {Nielsen}}, \ and\ \bibinfo {author} {\bibfnamefont {J.~R.}\ \bibnamefont {Petta}},\ }\href {\doibase 10.1126/sciadv.abn5130} {\bibfield  {journal} {\bibinfo  {journal} {Sci. Adv.}\ }\textbf {\bibinfo {volume} {8}},\ \bibinfo {pages} {eabn5130} (\bibinfo {year} {2022})}\BibitemShut {NoStop}%
\bibitem [{\citenamefont {Bennett}\ \emph {et~al.}(1992)\citenamefont {Bennett}, \citenamefont {Bessette}, \citenamefont {Brassard}, \citenamefont {Salvail},\ and\ \citenamefont {Smolin}}]{Bennett_JC_1992}%
  \BibitemOpen
  \bibfield  {author} {\bibinfo {author} {\bibfnamefont {C.~H.}\ \bibnamefont {Bennett}}, \bibinfo {author} {\bibfnamefont {F.}~\bibnamefont {Bessette}}, \bibinfo {author} {\bibfnamefont {G.}~\bibnamefont {Brassard}}, \bibinfo {author} {\bibfnamefont {L.}~\bibnamefont {Salvail}}, \ and\ \bibinfo {author} {\bibfnamefont {J.}~\bibnamefont {Smolin}},\ }\href {\doibase 10.1007/BF00191318} {\bibfield  {journal} {\bibinfo  {journal} {Journal of Cryptology}\ }\textbf {\bibinfo {volume} {5}},\ \bibinfo {pages} {3} (\bibinfo {year} {1992})}\BibitemShut {NoStop}%
\bibitem [{\citenamefont {Knill}\ \emph {et~al.}(2001)\citenamefont {Knill}, \citenamefont {Laflamme},\ and\ \citenamefont {Milburn}}]{Knill_Nature_2001}%
  \BibitemOpen
  \bibfield  {author} {\bibinfo {author} {\bibfnamefont {E.}~\bibnamefont {Knill}}, \bibinfo {author} {\bibfnamefont {R.}~\bibnamefont {Laflamme}}, \ and\ \bibinfo {author} {\bibfnamefont {G.~J.}\ \bibnamefont {Milburn}},\ }\href {\doibase 10.1038/35051009} {\bibfield  {journal} {\bibinfo  {journal} {Nature}\ }\textbf {\bibinfo {volume} {409}},\ \bibinfo {pages} {46} (\bibinfo {year} {2001})}\BibitemShut {NoStop}%
\bibitem [{\citenamefont {Henriet}\ \emph {et~al.}(2020)\citenamefont {Henriet}, \citenamefont {Beguin}, \citenamefont {Signoles}, \citenamefont {Lahaye}, \citenamefont {Browaeys}, \citenamefont {Reymond},\ and\ \citenamefont {Jurczak}}]{Henriet_Quantum_2020}%
  \BibitemOpen
  \bibfield  {author} {\bibinfo {author} {\bibfnamefont {L.}~\bibnamefont {Henriet}}, \bibinfo {author} {\bibfnamefont {L.}~\bibnamefont {Beguin}}, \bibinfo {author} {\bibfnamefont {A.}~\bibnamefont {Signoles}}, \bibinfo {author} {\bibfnamefont {T.}~\bibnamefont {Lahaye}}, \bibinfo {author} {\bibfnamefont {A.}~\bibnamefont {Browaeys}}, \bibinfo {author} {\bibfnamefont {G.-O.}\ \bibnamefont {Reymond}}, \ and\ \bibinfo {author} {\bibfnamefont {C.}~\bibnamefont {Jurczak}},\ }\href {\doibase 10.22331/q-2020-09-21-327} {\bibfield  {journal} {\bibinfo  {journal} {Quantum}\ }\textbf {\bibinfo {volume} {4}},\ \bibinfo {pages} {327} (\bibinfo {year} {2020})}\BibitemShut {NoStop}%
\bibitem [{\citenamefont {Bluvstein}\ \emph {et~al.}(2022)\citenamefont {Bluvstein}, \citenamefont {Levine}, \citenamefont {Semeghini}, \citenamefont {Wang}, \citenamefont {Ebadi}, \citenamefont {Kalinowski}, \citenamefont {Keesling}, \citenamefont {Maskara}, \citenamefont {Pichler}, \citenamefont {Greiner}, \citenamefont {Vuletić},\ and\ \citenamefont {Lukin}}]{Bluvstein_Nature_2022}%
  \BibitemOpen
  \bibfield  {author} {\bibinfo {author} {\bibfnamefont {D.}~\bibnamefont {Bluvstein}}, \bibinfo {author} {\bibfnamefont {H.}~\bibnamefont {Levine}}, \bibinfo {author} {\bibfnamefont {G.}~\bibnamefont {Semeghini}}, \bibinfo {author} {\bibfnamefont {T.~T.}\ \bibnamefont {Wang}}, \bibinfo {author} {\bibfnamefont {S.}~\bibnamefont {Ebadi}}, \bibinfo {author} {\bibfnamefont {M.}~\bibnamefont {Kalinowski}}, \bibinfo {author} {\bibfnamefont {A.}~\bibnamefont {Keesling}}, \bibinfo {author} {\bibfnamefont {N.}~\bibnamefont {Maskara}}, \bibinfo {author} {\bibfnamefont {H.}~\bibnamefont {Pichler}}, \bibinfo {author} {\bibfnamefont {M.}~\bibnamefont {Greiner}}, \bibinfo {author} {\bibfnamefont {V.}~\bibnamefont {Vuletić}}, \ and\ \bibinfo {author} {\bibfnamefont {M.~D.}\ \bibnamefont {Lukin}},\ }\href {\doibase 10.1038/s41586-022-04592-6} {\bibfield  {journal} {\bibinfo  {journal} {Nature}\ }\textbf {\bibinfo {volume} {604}},\ \bibinfo {pages} {451} (\bibinfo {year} {2022})}\BibitemShut {NoStop}%
\bibitem [{\citenamefont {Graham}\ \emph {et~al.}(2022)\citenamefont {Graham}, \citenamefont {Song}, \citenamefont {Scott}, \citenamefont {Poole}, \citenamefont {Phuttitarn}, \citenamefont {Jooya}, \citenamefont {Eichler}, \citenamefont {Jiang}, \citenamefont {Marra}, \citenamefont {Grinkemeyer}, \citenamefont {Kwon}, \citenamefont {Ebert}, \citenamefont {Cherek}, \citenamefont {Lichtman}, \citenamefont {Gillette}, \citenamefont {Gilbert}, \citenamefont {Bowman}, \citenamefont {Ballance}, \citenamefont {Campbell}, \citenamefont {Dahl}, \citenamefont {Crawford}, \citenamefont {Blunt}, \citenamefont {Rogers}, \citenamefont {Noel},\ and\ \citenamefont {Saffman}}]{Graham_Nature_2022}%
  \BibitemOpen
  \bibfield  {author} {\bibinfo {author} {\bibfnamefont {T.~M.}\ \bibnamefont {Graham}}, \bibinfo {author} {\bibfnamefont {Y.}~\bibnamefont {Song}}, \bibinfo {author} {\bibfnamefont {J.}~\bibnamefont {Scott}}, \bibinfo {author} {\bibfnamefont {C.}~\bibnamefont {Poole}}, \bibinfo {author} {\bibfnamefont {L.}~\bibnamefont {Phuttitarn}}, \bibinfo {author} {\bibfnamefont {K.}~\bibnamefont {Jooya}}, \bibinfo {author} {\bibfnamefont {P.}~\bibnamefont {Eichler}}, \bibinfo {author} {\bibfnamefont {X.}~\bibnamefont {Jiang}}, \bibinfo {author} {\bibfnamefont {A.}~\bibnamefont {Marra}}, \bibinfo {author} {\bibfnamefont {B.}~\bibnamefont {Grinkemeyer}}, \bibinfo {author} {\bibfnamefont {M.}~\bibnamefont {Kwon}}, \bibinfo {author} {\bibfnamefont {M.}~\bibnamefont {Ebert}}, \bibinfo {author} {\bibfnamefont {J.}~\bibnamefont {Cherek}}, \bibinfo {author} {\bibfnamefont {M.~T.}\ \bibnamefont {Lichtman}}, \bibinfo {author} {\bibfnamefont {M.}~\bibnamefont {Gillette}}, \bibinfo {author} {\bibfnamefont {J.}~\bibnamefont
  {Gilbert}}, \bibinfo {author} {\bibfnamefont {D.}~\bibnamefont {Bowman}}, \bibinfo {author} {\bibfnamefont {T.}~\bibnamefont {Ballance}}, \bibinfo {author} {\bibfnamefont {C.}~\bibnamefont {Campbell}}, \bibinfo {author} {\bibfnamefont {E.~D.}\ \bibnamefont {Dahl}}, \bibinfo {author} {\bibfnamefont {O.}~\bibnamefont {Crawford}}, \bibinfo {author} {\bibfnamefont {N.~S.}\ \bibnamefont {Blunt}}, \bibinfo {author} {\bibfnamefont {B.}~\bibnamefont {Rogers}}, \bibinfo {author} {\bibfnamefont {T.}~\bibnamefont {Noel}}, \ and\ \bibinfo {author} {\bibfnamefont {M.}~\bibnamefont {Saffman}},\ }\href {\doibase 10.1038/s41586-022-04603-6} {\bibfield  {journal} {\bibinfo  {journal} {Nature}\ }\textbf {\bibinfo {volume} {604}},\ \bibinfo {pages} {457} (\bibinfo {year} {2022})}\BibitemShut {NoStop}%
\bibitem [{\citenamefont {Park}\ \emph {et~al.}(2022)\citenamefont {Park}, \citenamefont {Trautmann}, \citenamefont {\ifmmode \check{S}\else \v{S}\fi{}anti\ifmmode~\acute{c}\else \'{c}\fi{}}, \citenamefont {Kl\"usener}, \citenamefont {Heinz}, \citenamefont {Bloch},\ and\ \citenamefont {Blatt}}]{Park_PRX_2022}%
  \BibitemOpen
  \bibfield  {author} {\bibinfo {author} {\bibfnamefont {A.~J.}\ \bibnamefont {Park}}, \bibinfo {author} {\bibfnamefont {J.}~\bibnamefont {Trautmann}}, \bibinfo {author} {\bibfnamefont {N.}~\bibnamefont {\ifmmode \check{S}\else \v{S}\fi{}anti\ifmmode~\acute{c}\else \'{c}\fi{}}}, \bibinfo {author} {\bibfnamefont {V.}~\bibnamefont {Kl\"usener}}, \bibinfo {author} {\bibfnamefont {A.}~\bibnamefont {Heinz}}, \bibinfo {author} {\bibfnamefont {I.}~\bibnamefont {Bloch}}, \ and\ \bibinfo {author} {\bibfnamefont {S.}~\bibnamefont {Blatt}},\ }\href {\doibase 10.1103/PRXQuantum.3.030314} {\bibfield  {journal} {\bibinfo  {journal} {PRX Quantum}\ }\textbf {\bibinfo {volume} {3}},\ \bibinfo {pages} {030314} (\bibinfo {year} {2022})}\BibitemShut {NoStop}%
\bibitem [{\citenamefont {Ferraro}\ \emph {et~al.}(2005)\citenamefont {Ferraro}, \citenamefont {Olivares},\ and\ \citenamefont {Paris}}]{Ferraro_Bib_2005}%
  \BibitemOpen
  \bibfield  {author} {\bibinfo {author} {\bibfnamefont {A.}~\bibnamefont {Ferraro}}, \bibinfo {author} {\bibfnamefont {S.}~\bibnamefont {Olivares}}, \ and\ \bibinfo {author} {\bibfnamefont {M.~G.~A.}\ \bibnamefont {Paris}},\ }\href {\doibase https://bibliopolis.it/napoli-series-on-physics-and-astrophysics/} {\emph {\bibinfo {title} {Gaussian States in Quantum Information}}}\ (\bibinfo  {publisher} {Bibliopolis, Napoli},\ \bibinfo {year} {2005})\BibitemShut {NoStop}%
\bibitem [{\citenamefont {Adesso}\ \emph {et~al.}(2014)\citenamefont {Adesso}, \citenamefont {Ragy},\ and\ \citenamefont {Lee}}]{Adesso_OSID_2014}%
  \BibitemOpen
  \bibfield  {author} {\bibinfo {author} {\bibfnamefont {G.}~\bibnamefont {Adesso}}, \bibinfo {author} {\bibfnamefont {S.}~\bibnamefont {Ragy}}, \ and\ \bibinfo {author} {\bibfnamefont {A.~R.}\ \bibnamefont {Lee}},\ }\href {\doibase 10.1142/S1230161214400010} {\bibfield  {journal} {\bibinfo  {journal} {Open Systems and Information Dynamics}\ }\textbf {\bibinfo {volume} {21}},\ \bibinfo {pages} {1} (\bibinfo {year} {2014})}\BibitemShut {NoStop}%
\bibitem [{\citenamefont {Serafini}(2017)}]{Serafini_2017}%
  \BibitemOpen
  \bibfield  {author} {\bibinfo {author} {\bibfnamefont {A.}~\bibnamefont {Serafini}},\ }\href {\doibase https://doi.org/10.1201/9781315118727} {\emph {\bibinfo {title} {Quantum Continuous Variables}}},\ \bibinfo {edition} {1st}\ ed.\ (\bibinfo  {publisher} {CRC Press},\ \bibinfo {year} {2017})\BibitemShut {NoStop}%
\bibitem [{\citenamefont {Walls}\ and\ \citenamefont {Milburn}(1994)}]{Walls-Millburn_1994}%
  \BibitemOpen
  \bibfield  {author} {\bibinfo {author} {\bibfnamefont {D.~F.}\ \bibnamefont {Walls}}\ and\ \bibinfo {author} {\bibfnamefont {G.~J.}\ \bibnamefont {Milburn}},\ }\href {\doibase 10.1007/978-3-642-79504-6} {\emph {\bibinfo {title} {Quantum Optics}}}\ (\bibinfo  {publisher} {Springer Berlin Heidelberg},\ \bibinfo {year} {1994})\BibitemShut {NoStop}%
\bibitem [{\citenamefont {Braunstein}\ and\ \citenamefont {Loock}(2005)}]{Braunstein_RMP_2005}%
  \BibitemOpen
  \bibfield  {author} {\bibinfo {author} {\bibfnamefont {L.~S.}\ \bibnamefont {Braunstein}}\ and\ \bibinfo {author} {\bibfnamefont {P.~V.}\ \bibnamefont {Loock}},\ }\href {\doibase 10.1103/RevModPhys.77.513} {\bibfield  {journal} {\bibinfo  {journal} {Reviews of Modern Physics}\ }\textbf {\bibinfo {volume} {77}},\ \bibinfo {pages} {513} (\bibinfo {year} {2005})}\BibitemShut {NoStop}%
\bibitem [{\citenamefont {Reck}\ \emph {et~al.}(1994)\citenamefont {Reck}, \citenamefont {Zeilinger}, \citenamefont {Bernstein},\ and\ \citenamefont {Bertani}}]{Reck_PRL_1994}%
  \BibitemOpen
  \bibfield  {author} {\bibinfo {author} {\bibfnamefont {M.}~\bibnamefont {Reck}}, \bibinfo {author} {\bibfnamefont {A.}~\bibnamefont {Zeilinger}}, \bibinfo {author} {\bibfnamefont {H.~J.}\ \bibnamefont {Bernstein}}, \ and\ \bibinfo {author} {\bibfnamefont {P.}~\bibnamefont {Bertani}},\ }\href {\doibase 10.1103/PhysRevLett.73.58} {\bibfield  {journal} {\bibinfo  {journal} {Phys. Rev. Lett.}\ }\textbf {\bibinfo {volume} {73}},\ \bibinfo {pages} {58} (\bibinfo {year} {1994})}\BibitemShut {NoStop}%
\bibitem [{\citenamefont {Calsamiglia}\ and\ \citenamefont {Lütkenhaus}(2001)}]{Calsamiglia_APB_2001}%
  \BibitemOpen
  \bibfield  {author} {\bibinfo {author} {\bibfnamefont {J.}~\bibnamefont {Calsamiglia}}\ and\ \bibinfo {author} {\bibfnamefont {N.}~\bibnamefont {Lütkenhaus}},\ }\href {\doibase 10.1007/s003400000484} {\bibfield  {journal} {\bibinfo  {journal} {Applied Physics B}\ }\textbf {\bibinfo {volume} {72}},\ \bibinfo {pages} {67} (\bibinfo {year} {2001})}\BibitemShut {NoStop}%
\bibitem [{\citenamefont {Chabaud}\ \emph {et~al.}(2017)\citenamefont {Chabaud}, \citenamefont {Douce}, \citenamefont {Markham}, \citenamefont {van Loock}, \citenamefont {Kashefi},\ and\ \citenamefont {Ferrini}}]{Chabaud_PRA_2017}%
  \BibitemOpen
  \bibfield  {author} {\bibinfo {author} {\bibfnamefont {U.}~\bibnamefont {Chabaud}}, \bibinfo {author} {\bibfnamefont {T.}~\bibnamefont {Douce}}, \bibinfo {author} {\bibfnamefont {D.}~\bibnamefont {Markham}}, \bibinfo {author} {\bibfnamefont {P.}~\bibnamefont {van Loock}}, \bibinfo {author} {\bibfnamefont {E.}~\bibnamefont {Kashefi}}, \ and\ \bibinfo {author} {\bibfnamefont {G.}~\bibnamefont {Ferrini}},\ }\href {\doibase 10.1103/PhysRevA.96.062307} {\bibfield  {journal} {\bibinfo  {journal} {Physical Review A}\ }\textbf {\bibinfo {volume} {96}},\ \bibinfo {pages} {062307} (\bibinfo {year} {2017})}\BibitemShut {NoStop}%
\bibitem [{\citenamefont {Eisert}\ \emph {et~al.}(2020)\citenamefont {Eisert}, \citenamefont {Hangleiter}, \citenamefont {Walk}, \citenamefont {Roth}, \citenamefont {Markham}, \citenamefont {Parekh}, \citenamefont {Chabaud},\ and\ \citenamefont {Kashefi}}]{Eisert_Nature_2020}%
  \BibitemOpen
  \bibfield  {author} {\bibinfo {author} {\bibfnamefont {J.}~\bibnamefont {Eisert}}, \bibinfo {author} {\bibfnamefont {D.}~\bibnamefont {Hangleiter}}, \bibinfo {author} {\bibfnamefont {N.}~\bibnamefont {Walk}}, \bibinfo {author} {\bibfnamefont {I.}~\bibnamefont {Roth}}, \bibinfo {author} {\bibfnamefont {D.}~\bibnamefont {Markham}}, \bibinfo {author} {\bibfnamefont {R.}~\bibnamefont {Parekh}}, \bibinfo {author} {\bibfnamefont {U.}~\bibnamefont {Chabaud}}, \ and\ \bibinfo {author} {\bibfnamefont {E.}~\bibnamefont {Kashefi}},\ }\href {\doibase 10.1038/s42254-020-0186-4} {\bibfield  {journal} {\bibinfo  {journal} {Nature Reviews Physics}\ }\textbf {\bibinfo {volume} {2}},\ \bibinfo {pages} {382} (\bibinfo {year} {2020})}\BibitemShut {NoStop}%
\bibitem [{\citenamefont {Chabaud}\ \emph {et~al.}(2021{\natexlab{a}})\citenamefont {Chabaud}, \citenamefont {Ferrini}, \citenamefont {Grosshans},\ and\ \citenamefont {Markham}}]{Chabaud_PRR_2021}%
  \BibitemOpen
  \bibfield  {author} {\bibinfo {author} {\bibfnamefont {U.}~\bibnamefont {Chabaud}}, \bibinfo {author} {\bibfnamefont {G.}~\bibnamefont {Ferrini}}, \bibinfo {author} {\bibfnamefont {F.}~\bibnamefont {Grosshans}}, \ and\ \bibinfo {author} {\bibfnamefont {D.}~\bibnamefont {Markham}},\ }\href {\doibase 10.1103/PhysRevResearch.3.033018} {\bibfield  {journal} {\bibinfo  {journal} {Phys. Rev. Res.}\ }\textbf {\bibinfo {volume} {3}},\ \bibinfo {pages} {033018} (\bibinfo {year} {2021}{\natexlab{a}})}\BibitemShut {NoStop}%
\bibitem [{\citenamefont {Chabaud}\ \emph {et~al.}(2021{\natexlab{b}})\citenamefont {Chabaud}, \citenamefont {Grosshans}, \citenamefont {Kashefi},\ and\ \citenamefont {Markham}}]{Chabaud_Quantum_2021}%
  \BibitemOpen
  \bibfield  {author} {\bibinfo {author} {\bibfnamefont {U.}~\bibnamefont {Chabaud}}, \bibinfo {author} {\bibfnamefont {F.}~\bibnamefont {Grosshans}}, \bibinfo {author} {\bibfnamefont {E.}~\bibnamefont {Kashefi}}, \ and\ \bibinfo {author} {\bibfnamefont {D.}~\bibnamefont {Markham}},\ }\href {\doibase 10.22331/q-2021-11-15-578} {\bibfield  {journal} {\bibinfo  {journal} {Quantum}\ }\textbf {\bibinfo {volume} {5}},\ \bibinfo {pages} {578} (\bibinfo {year} {2021}{\natexlab{b}})}\BibitemShut {NoStop}%
\bibitem [{\citenamefont {Calcluth}\ \emph {et~al.}(2022)\citenamefont {Calcluth}, \citenamefont {Ferraro},\ and\ \citenamefont {Ferrini}}]{Calcluth_Quantum_2022}%
  \BibitemOpen
  \bibfield  {author} {\bibinfo {author} {\bibfnamefont {C.}~\bibnamefont {Calcluth}}, \bibinfo {author} {\bibfnamefont {A.}~\bibnamefont {Ferraro}}, \ and\ \bibinfo {author} {\bibfnamefont {G.}~\bibnamefont {Ferrini}},\ }\href {\doibase 10.22331/q-2022-12-01-867} {\bibfield  {journal} {\bibinfo  {journal} {Quantum}\ }\textbf {\bibinfo {volume} {6}},\ \bibinfo {pages} {867} (\bibinfo {year} {2022})}\BibitemShut {NoStop}%
\bibitem [{\citenamefont {Calcluth}\ \emph {et~al.}(2023)\citenamefont {Calcluth}, \citenamefont {Ferraro},\ and\ \citenamefont {Ferrini}}]{Calcluth_PRA_2023}%
  \BibitemOpen
  \bibfield  {author} {\bibinfo {author} {\bibfnamefont {C.}~\bibnamefont {Calcluth}}, \bibinfo {author} {\bibfnamefont {A.}~\bibnamefont {Ferraro}}, \ and\ \bibinfo {author} {\bibfnamefont {G.}~\bibnamefont {Ferrini}},\ }\href {\doibase 10.1103/PhysRevA.107.062414} {\bibfield  {journal} {\bibinfo  {journal} {Phys. Rev. A}\ }\textbf {\bibinfo {volume} {107}},\ \bibinfo {pages} {062414} (\bibinfo {year} {2023})}\BibitemShut {NoStop}%
\bibitem [{\citenamefont {Calcluth}\ \emph {et~al.}(2024)\citenamefont {Calcluth}, \citenamefont {Reichel}, \citenamefont {Ferraro},\ and\ \citenamefont {Ferrini}}]{Calcluth_PRX_2024}%
  \BibitemOpen
  \bibfield  {author} {\bibinfo {author} {\bibfnamefont {C.}~\bibnamefont {Calcluth}}, \bibinfo {author} {\bibfnamefont {N.}~\bibnamefont {Reichel}}, \bibinfo {author} {\bibfnamefont {A.}~\bibnamefont {Ferraro}}, \ and\ \bibinfo {author} {\bibfnamefont {G.}~\bibnamefont {Ferrini}},\ }\href {\doibase 10.1103/PRXQuantum.5.020337} {\bibfield  {journal} {\bibinfo  {journal} {PRX Quantum}\ }\textbf {\bibinfo {volume} {5}},\ \bibinfo {pages} {020337} (\bibinfo {year} {2024})}\BibitemShut {NoStop}%
\bibitem [{\citenamefont {Lloyd}\ and\ \citenamefont {Braunstein}(1999)}]{LLoyd_PRL_1999}%
  \BibitemOpen
  \bibfield  {author} {\bibinfo {author} {\bibfnamefont {S.}~\bibnamefont {Lloyd}}\ and\ \bibinfo {author} {\bibfnamefont {S.~L.}\ \bibnamefont {Braunstein}},\ }\href {\doibase 10.1103/PhysRevLett.82.1784} {\bibfield  {journal} {\bibinfo  {journal} {Phys. Rev. Lett.}\ }\textbf {\bibinfo {volume} {82}},\ \bibinfo {pages} {1784} (\bibinfo {year} {1999})}\BibitemShut {NoStop}%
\bibitem [{\citenamefont {Houhou}\ \emph {et~al.}(2022)\citenamefont {Houhou}, \citenamefont {Moore}, \citenamefont {Bose},\ and\ \citenamefont {Ferraro}}]{Houhou_PRA_2022}%
  \BibitemOpen
  \bibfield  {author} {\bibinfo {author} {\bibfnamefont {O.}~\bibnamefont {Houhou}}, \bibinfo {author} {\bibfnamefont {D.~W.}\ \bibnamefont {Moore}}, \bibinfo {author} {\bibfnamefont {S.}~\bibnamefont {Bose}}, \ and\ \bibinfo {author} {\bibfnamefont {A.}~\bibnamefont {Ferraro}},\ }\href {\doibase 10.1103/PhysRevA.105.012610} {\bibfield  {journal} {\bibinfo  {journal} {Phys. Rev. A}\ }\textbf {\bibinfo {volume} {105}},\ \bibinfo {pages} {012610} (\bibinfo {year} {2022})}\BibitemShut {NoStop}%
\bibitem [{\citenamefont {Kudra}\ \emph {et~al.}(2022)\citenamefont {Kudra}, \citenamefont {Kervinen}, \citenamefont {Strandberg}, \citenamefont {Ahmed}, \citenamefont {Scigliuzzo}, \citenamefont {Osman}, \citenamefont {Lozano}, \citenamefont {Thol\'en}, \citenamefont {Borgani}, \citenamefont {Haviland}, \citenamefont {Ferrini}, \citenamefont {Bylander}, \citenamefont {Kockum}, \citenamefont {Quijandr\'{\i}a}, \citenamefont {Delsing},\ and\ \citenamefont {Gasparinetti}}]{Kundra_PRX_2022}%
  \BibitemOpen
  \bibfield  {author} {\bibinfo {author} {\bibfnamefont {M.}~\bibnamefont {Kudra}}, \bibinfo {author} {\bibfnamefont {M.}~\bibnamefont {Kervinen}}, \bibinfo {author} {\bibfnamefont {I.}~\bibnamefont {Strandberg}}, \bibinfo {author} {\bibfnamefont {S.}~\bibnamefont {Ahmed}}, \bibinfo {author} {\bibfnamefont {M.}~\bibnamefont {Scigliuzzo}}, \bibinfo {author} {\bibfnamefont {A.}~\bibnamefont {Osman}}, \bibinfo {author} {\bibfnamefont {D.~P.}\ \bibnamefont {Lozano}}, \bibinfo {author} {\bibfnamefont {M.~O.}\ \bibnamefont {Thol\'en}}, \bibinfo {author} {\bibfnamefont {R.}~\bibnamefont {Borgani}}, \bibinfo {author} {\bibfnamefont {D.~B.}\ \bibnamefont {Haviland}}, \bibinfo {author} {\bibfnamefont {G.}~\bibnamefont {Ferrini}}, \bibinfo {author} {\bibfnamefont {J.}~\bibnamefont {Bylander}}, \bibinfo {author} {\bibfnamefont {A.~F.}\ \bibnamefont {Kockum}}, \bibinfo {author} {\bibfnamefont {F.}~\bibnamefont {Quijandr\'{\i}a}}, \bibinfo {author} {\bibfnamefont {P.}~\bibnamefont {Delsing}}, \ and\ \bibinfo {author}
  {\bibfnamefont {S.}~\bibnamefont {Gasparinetti}},\ }\href {\doibase 10.1103/PRXQuantum.3.030301} {\bibfield  {journal} {\bibinfo  {journal} {PRX Quantum}\ }\textbf {\bibinfo {volume} {3}},\ \bibinfo {pages} {030301} (\bibinfo {year} {2022})}\BibitemShut {NoStop}%
\bibitem [{\citenamefont {Zheng}\ \emph {et~al.}(2023)\citenamefont {Zheng}, \citenamefont {Ferraro}, \citenamefont {Kockum},\ and\ \citenamefont {Ferrini}}]{Zheng_PRA_2023}%
  \BibitemOpen
  \bibfield  {author} {\bibinfo {author} {\bibfnamefont {Y.}~\bibnamefont {Zheng}}, \bibinfo {author} {\bibfnamefont {A.}~\bibnamefont {Ferraro}}, \bibinfo {author} {\bibfnamefont {A.~F.}\ \bibnamefont {Kockum}}, \ and\ \bibinfo {author} {\bibfnamefont {G.}~\bibnamefont {Ferrini}},\ }\href {\doibase 10.1103/PhysRevA.108.012603} {\bibfield  {journal} {\bibinfo  {journal} {Phys. Rev. A}\ }\textbf {\bibinfo {volume} {108}},\ \bibinfo {pages} {012603} (\bibinfo {year} {2023})}\BibitemShut {NoStop}%
\bibitem [{\citenamefont {McConnell}\ \emph {et~al.}(2024)\citenamefont {McConnell}, \citenamefont {Houhou}, \citenamefont {Brunelli},\ and\ \citenamefont {Ferraro}}]{McConnell_PRA_2024}%
  \BibitemOpen
  \bibfield  {author} {\bibinfo {author} {\bibfnamefont {P.}~\bibnamefont {McConnell}}, \bibinfo {author} {\bibfnamefont {O.}~\bibnamefont {Houhou}}, \bibinfo {author} {\bibfnamefont {M.}~\bibnamefont {Brunelli}}, \ and\ \bibinfo {author} {\bibfnamefont {A.}~\bibnamefont {Ferraro}},\ }\href {\doibase 10.1103/PhysRevA.109.033508} {\bibfield  {journal} {\bibinfo  {journal} {Phys. Rev. A}\ }\textbf {\bibinfo {volume} {109}},\ \bibinfo {pages} {033508} (\bibinfo {year} {2024})}\BibitemShut {NoStop}%
\bibitem [{\citenamefont {Hillmann}\ \emph {et~al.}(2020)\citenamefont {Hillmann}, \citenamefont {Quijandr\'{\i}a}, \citenamefont {Johansson}, \citenamefont {Ferraro}, \citenamefont {Gasparinetti},\ and\ \citenamefont {Ferrini}}]{Hillman_PRL_2020}%
  \BibitemOpen
  \bibfield  {author} {\bibinfo {author} {\bibfnamefont {T.}~\bibnamefont {Hillmann}}, \bibinfo {author} {\bibfnamefont {F.}~\bibnamefont {Quijandr\'{\i}a}}, \bibinfo {author} {\bibfnamefont {G.}~\bibnamefont {Johansson}}, \bibinfo {author} {\bibfnamefont {A.}~\bibnamefont {Ferraro}}, \bibinfo {author} {\bibfnamefont {S.}~\bibnamefont {Gasparinetti}}, \ and\ \bibinfo {author} {\bibfnamefont {G.}~\bibnamefont {Ferrini}},\ }\href {\doibase 10.1103/PhysRevLett.125.160501} {\bibfield  {journal} {\bibinfo  {journal} {Phys. Rev. Lett.}\ }\textbf {\bibinfo {volume} {125}},\ \bibinfo {pages} {160501} (\bibinfo {year} {2020})}\BibitemShut {NoStop}%
\bibitem [{\citenamefont {Chapman}\ \emph {et~al.}(2023)\citenamefont {Chapman}, \citenamefont {de~Graaf}, \citenamefont {Xue}, \citenamefont {Zhang}, \citenamefont {Teoh}, \citenamefont {Curtis}, \citenamefont {Tsunoda}, \citenamefont {Eickbusch}, \citenamefont {Read}, \citenamefont {Koottandavida}, \citenamefont {Mundhada}, \citenamefont {Frunzio}, \citenamefont {Devoret}, \citenamefont {Girvin},\ and\ \citenamefont {Schoelkopf}}]{Chapman_PRX_2023}%
  \BibitemOpen
  \bibfield  {author} {\bibinfo {author} {\bibfnamefont {B.~J.}\ \bibnamefont {Chapman}}, \bibinfo {author} {\bibfnamefont {S.~J.}\ \bibnamefont {de~Graaf}}, \bibinfo {author} {\bibfnamefont {S.~H.}\ \bibnamefont {Xue}}, \bibinfo {author} {\bibfnamefont {Y.}~\bibnamefont {Zhang}}, \bibinfo {author} {\bibfnamefont {J.}~\bibnamefont {Teoh}}, \bibinfo {author} {\bibfnamefont {J.~C.}\ \bibnamefont {Curtis}}, \bibinfo {author} {\bibfnamefont {T.}~\bibnamefont {Tsunoda}}, \bibinfo {author} {\bibfnamefont {A.}~\bibnamefont {Eickbusch}}, \bibinfo {author} {\bibfnamefont {A.~P.}\ \bibnamefont {Read}}, \bibinfo {author} {\bibfnamefont {A.}~\bibnamefont {Koottandavida}}, \bibinfo {author} {\bibfnamefont {S.~O.}\ \bibnamefont {Mundhada}}, \bibinfo {author} {\bibfnamefont {L.}~\bibnamefont {Frunzio}}, \bibinfo {author} {\bibfnamefont {M.}~\bibnamefont {Devoret}}, \bibinfo {author} {\bibfnamefont {S.}~\bibnamefont {Girvin}}, \ and\ \bibinfo {author} {\bibfnamefont {R.}~\bibnamefont {Schoelkopf}},\ }\href {\doibase
  10.1103/PRXQuantum.4.020355} {\bibfield  {journal} {\bibinfo  {journal} {PRX Quantum}\ }\textbf {\bibinfo {volume} {4}},\ \bibinfo {pages} {020355} (\bibinfo {year} {2023})}\BibitemShut {NoStop}%
\bibitem [{\citenamefont {Zhou}\ \emph {et~al.}(2023)\citenamefont {Zhou}, \citenamefont {Lu}, \citenamefont {Praquin}, \citenamefont {Chien}, \citenamefont {Kaufman}, \citenamefont {Cao}, \citenamefont {Xia}, \citenamefont {Mong}, \citenamefont {Pfaff}, \citenamefont {Pekker},\ and\ \citenamefont {Hatridge}}]{Zhou_NPJ_2023}%
  \BibitemOpen
  \bibfield  {author} {\bibinfo {author} {\bibfnamefont {C.}~\bibnamefont {Zhou}}, \bibinfo {author} {\bibfnamefont {P.}~\bibnamefont {Lu}}, \bibinfo {author} {\bibfnamefont {M.}~\bibnamefont {Praquin}}, \bibinfo {author} {\bibfnamefont {T.-C.}\ \bibnamefont {Chien}}, \bibinfo {author} {\bibfnamefont {R.}~\bibnamefont {Kaufman}}, \bibinfo {author} {\bibfnamefont {X.}~\bibnamefont {Cao}}, \bibinfo {author} {\bibfnamefont {M.}~\bibnamefont {Xia}}, \bibinfo {author} {\bibfnamefont {R.~S.~K.}\ \bibnamefont {Mong}}, \bibinfo {author} {\bibfnamefont {W.}~\bibnamefont {Pfaff}}, \bibinfo {author} {\bibfnamefont {D.}~\bibnamefont {Pekker}}, \ and\ \bibinfo {author} {\bibfnamefont {M.}~\bibnamefont {Hatridge}},\ }\href {\doibase 10.1038/s41534-023-00723-7} {\bibfield  {journal} {\bibinfo  {journal} {npj Quantum Information}\ }\textbf {\bibinfo {volume} {9}},\ \bibinfo {pages} {54} (\bibinfo {year} {2023})}\BibitemShut {NoStop}%
\bibitem [{\citenamefont {Strandberg}\ \emph {et~al.}(2024)\citenamefont {Strandberg}, \citenamefont {Eriksson}, \citenamefont {Royer}, \citenamefont {Kervinen},\ and\ \citenamefont {Gasparinetti}}]{Strandberg_PRL_2024}%
  \BibitemOpen
  \bibfield  {author} {\bibinfo {author} {\bibfnamefont {I.}~\bibnamefont {Strandberg}}, \bibinfo {author} {\bibfnamefont {A.~M.}\ \bibnamefont {Eriksson}}, \bibinfo {author} {\bibfnamefont {B.}~\bibnamefont {Royer}}, \bibinfo {author} {\bibfnamefont {M.}~\bibnamefont {Kervinen}}, \ and\ \bibinfo {author} {\bibfnamefont {S.}~\bibnamefont {Gasparinetti}},\ }\href {\doibase 10.1103/PhysRevLett.133.063601} {\bibfield  {journal} {\bibinfo  {journal} {Phys. Rev. Lett.}\ }\textbf {\bibinfo {volume} {133}},\ \bibinfo {pages} {063601} (\bibinfo {year} {2024})}\BibitemShut {NoStop}%
\bibitem [{\citenamefont {Eriksson}\ \emph {et~al.}(2024)\citenamefont {Eriksson}, \citenamefont {Sépulcre}, \citenamefont {Kervinen}, \citenamefont {Hillmann}, \citenamefont {Kudra}, \citenamefont {Dupouy}, \citenamefont {Lu}, \citenamefont {Khanahmadi}, \citenamefont {Yang}, \citenamefont {Castillo-Moreno}, \citenamefont {Delsing},\ and\ \citenamefont {Gasparinetti}}]{Eriksson_NC_2024}%
  \BibitemOpen
  \bibfield  {author} {\bibinfo {author} {\bibfnamefont {A.~M.}\ \bibnamefont {Eriksson}}, \bibinfo {author} {\bibfnamefont {T.}~\bibnamefont {Sépulcre}}, \bibinfo {author} {\bibfnamefont {M.}~\bibnamefont {Kervinen}}, \bibinfo {author} {\bibfnamefont {T.}~\bibnamefont {Hillmann}}, \bibinfo {author} {\bibfnamefont {M.}~\bibnamefont {Kudra}}, \bibinfo {author} {\bibfnamefont {S.}~\bibnamefont {Dupouy}}, \bibinfo {author} {\bibfnamefont {Y.}~\bibnamefont {Lu}}, \bibinfo {author} {\bibfnamefont {M.}~\bibnamefont {Khanahmadi}}, \bibinfo {author} {\bibfnamefont {J.}~\bibnamefont {Yang}}, \bibinfo {author} {\bibfnamefont {C.}~\bibnamefont {Castillo-Moreno}}, \bibinfo {author} {\bibfnamefont {P.}~\bibnamefont {Delsing}}, \ and\ \bibinfo {author} {\bibfnamefont {S.}~\bibnamefont {Gasparinetti}},\ }\href {\doibase 10.1038/s41467-024-46507-1} {\bibfield  {journal} {\bibinfo  {journal} {Nature Communications}\ }\textbf {\bibinfo {volume} {15}},\ \bibinfo {pages} {2512} (\bibinfo {year} {2024})}\BibitemShut {NoStop}%
\bibitem [{\citenamefont {Menicucci}\ \emph {et~al.}(2006)\citenamefont {Menicucci}, \citenamefont {van Loock}, \citenamefont {Gu}, \citenamefont {Weedbrook}, \citenamefont {Ralph},\ and\ \citenamefont {Nielsen}}]{Menicucci_PRL_2006}%
  \BibitemOpen
  \bibfield  {author} {\bibinfo {author} {\bibfnamefont {N.~C.}\ \bibnamefont {Menicucci}}, \bibinfo {author} {\bibfnamefont {P.}~\bibnamefont {van Loock}}, \bibinfo {author} {\bibfnamefont {M.}~\bibnamefont {Gu}}, \bibinfo {author} {\bibfnamefont {C.}~\bibnamefont {Weedbrook}}, \bibinfo {author} {\bibfnamefont {T.~C.}\ \bibnamefont {Ralph}}, \ and\ \bibinfo {author} {\bibfnamefont {M.~A.}\ \bibnamefont {Nielsen}},\ }\href {\doibase 10.1103/PhysRevLett.97.110501} {\bibfield  {journal} {\bibinfo  {journal} {Phys. Rev. Lett.}\ }\textbf {\bibinfo {volume} {97}},\ \bibinfo {pages} {110501} (\bibinfo {year} {2006})}\BibitemShut {NoStop}%
\bibitem [{\citenamefont {Breuer}\ and\ \citenamefont {Petruccione}(2007)}]{Breuer_2007}%
  \BibitemOpen
  \bibfield  {author} {\bibinfo {author} {\bibfnamefont {H.-P.}\ \bibnamefont {Breuer}}\ and\ \bibinfo {author} {\bibfnamefont {F.}~\bibnamefont {Petruccione}},\ }\href {\doibase 10.1093/acprof:oso/9780199213900.001.0001} {\emph {\bibinfo {title} {The Theory of Open Quantum Systems}}}\ (\bibinfo  {publisher} {Oxford University Press},\ \bibinfo {year} {2007})\BibitemShut {NoStop}%
\bibitem [{\citenamefont {Lidar}(2019)}]{Lidar_arXiv_2019}%
  \BibitemOpen
  \bibfield  {author} {\bibinfo {author} {\bibfnamefont {D.~A.}\ \bibnamefont {Lidar}},\ }\href {https://doi.org/10.48550/arXiv.1902.00967} {\bibfield  {journal} {\bibinfo  {journal} {arXiv:1902.00967}\ } (\bibinfo {year} {2019})}\BibitemShut {NoStop}%
\bibitem [{\citenamefont {Rivas}\ and\ \citenamefont {Huelga}(2012)}]{Rivas_2012}%
  \BibitemOpen
  \bibfield  {author} {\bibinfo {author} {\bibfnamefont {A.}~\bibnamefont {Rivas}}\ and\ \bibinfo {author} {\bibfnamefont {S.~F.}\ \bibnamefont {Huelga}},\ }\href {\doibase 10.1007/978-3-642-23354-8} {\emph {\bibinfo {title} {Open Quantum Systems}}}\ (\bibinfo  {publisher} {Springer Berlin Heidelberg},\ \bibinfo {year} {2012})\BibitemShut {NoStop}%
\bibitem [{\citenamefont {Silva}\ \emph {et~al.}(2015)\citenamefont {Silva}, \citenamefont {Gisin}, \citenamefont {Guryanova},\ and\ \citenamefont {Popescu}}]{Silva_PRL_2015}%
  \BibitemOpen
  \bibfield  {author} {\bibinfo {author} {\bibfnamefont {R.}~\bibnamefont {Silva}}, \bibinfo {author} {\bibfnamefont {N.}~\bibnamefont {Gisin}}, \bibinfo {author} {\bibfnamefont {Y.}~\bibnamefont {Guryanova}}, \ and\ \bibinfo {author} {\bibfnamefont {S.}~\bibnamefont {Popescu}},\ }\href {\doibase 10.1103/PhysRevLett.114.250401} {\bibfield  {journal} {\bibinfo  {journal} {Phys. Rev. Lett.}\ }\textbf {\bibinfo {volume} {114}},\ \bibinfo {pages} {250401} (\bibinfo {year} {2015})}\BibitemShut {NoStop}%
\bibitem [{\citenamefont {Brunner}\ \emph {et~al.}(2014)\citenamefont {Brunner}, \citenamefont {Cavalcanti}, \citenamefont {Pironio}, \citenamefont {Scarani},\ and\ \citenamefont {Wehner}}]{Brunner_RMP_2014}%
  \BibitemOpen
  \bibfield  {author} {\bibinfo {author} {\bibfnamefont {N.}~\bibnamefont {Brunner}}, \bibinfo {author} {\bibfnamefont {D.}~\bibnamefont {Cavalcanti}}, \bibinfo {author} {\bibfnamefont {S.}~\bibnamefont {Pironio}}, \bibinfo {author} {\bibfnamefont {V.}~\bibnamefont {Scarani}}, \ and\ \bibinfo {author} {\bibfnamefont {S.}~\bibnamefont {Wehner}},\ }\href {\doibase 10.1103/RevModPhys.86.419} {\bibfield  {journal} {\bibinfo  {journal} {Reviews of Modern Physics}\ }\textbf {\bibinfo {volume} {86}},\ \bibinfo {pages} {419} (\bibinfo {year} {2014})}\BibitemShut {NoStop}%
\bibitem [{\citenamefont {Mal}\ \emph {et~al.}(2016)\citenamefont {Mal}, \citenamefont {Majumdar},\ and\ \citenamefont {Home}}]{Mal_Maths_2016}%
  \BibitemOpen
  \bibfield  {author} {\bibinfo {author} {\bibfnamefont {S.}~\bibnamefont {Mal}}, \bibinfo {author} {\bibfnamefont {A.}~\bibnamefont {Majumdar}}, \ and\ \bibinfo {author} {\bibfnamefont {D.}~\bibnamefont {Home}},\ }\href {\doibase 10.3390/math4030048} {\bibfield  {journal} {\bibinfo  {journal} {Mathematics}\ }\textbf {\bibinfo {volume} {4}},\ \bibinfo {pages} {48} (\bibinfo {year} {2016})}\BibitemShut {NoStop}%
\bibitem [{\citenamefont {Bera}\ \emph {et~al.}(2018)\citenamefont {Bera}, \citenamefont {Mal}, \citenamefont {Sen(De)},\ and\ \citenamefont {Sen}}]{Bera_PRA_2018}%
  \BibitemOpen
  \bibfield  {author} {\bibinfo {author} {\bibfnamefont {A.}~\bibnamefont {Bera}}, \bibinfo {author} {\bibfnamefont {S.}~\bibnamefont {Mal}}, \bibinfo {author} {\bibfnamefont {A.}~\bibnamefont {Sen(De)}}, \ and\ \bibinfo {author} {\bibfnamefont {U.}~\bibnamefont {Sen}},\ }\href {\doibase 10.1103/PhysRevA.98.062304} {\bibfield  {journal} {\bibinfo  {journal} {Phys. Rev. A}\ }\textbf {\bibinfo {volume} {98}},\ \bibinfo {pages} {062304} (\bibinfo {year} {2018})}\BibitemShut {NoStop}%
\bibitem [{\citenamefont {Sasmal}\ \emph {et~al.}(2018)\citenamefont {Sasmal}, \citenamefont {Das}, \citenamefont {Mal},\ and\ \citenamefont {Majumdar}}]{Sasmal_PRA_2018}%
  \BibitemOpen
  \bibfield  {author} {\bibinfo {author} {\bibfnamefont {S.}~\bibnamefont {Sasmal}}, \bibinfo {author} {\bibfnamefont {D.}~\bibnamefont {Das}}, \bibinfo {author} {\bibfnamefont {S.}~\bibnamefont {Mal}}, \ and\ \bibinfo {author} {\bibfnamefont {A.~S.}\ \bibnamefont {Majumdar}},\ }\href {\doibase 10.1103/PhysRevA.98.012305} {\bibfield  {journal} {\bibinfo  {journal} {Phys. Rev. A}\ }\textbf {\bibinfo {volume} {98}},\ \bibinfo {pages} {012305} (\bibinfo {year} {2018})}\BibitemShut {NoStop}%
\bibitem [{\citenamefont {Shenoy~H.}\ \emph {et~al.}(2019)\citenamefont {Shenoy~H.}, \citenamefont {Designolle}, \citenamefont {Hirsch}, \citenamefont {Silva}, \citenamefont {Gisin},\ and\ \citenamefont {Brunner}}]{Shenoy_PRA_2019}%
  \BibitemOpen
  \bibfield  {author} {\bibinfo {author} {\bibfnamefont {A.}~\bibnamefont {Shenoy~H.}}, \bibinfo {author} {\bibfnamefont {S.}~\bibnamefont {Designolle}}, \bibinfo {author} {\bibfnamefont {F.}~\bibnamefont {Hirsch}}, \bibinfo {author} {\bibfnamefont {R.}~\bibnamefont {Silva}}, \bibinfo {author} {\bibfnamefont {N.}~\bibnamefont {Gisin}}, \ and\ \bibinfo {author} {\bibfnamefont {N.}~\bibnamefont {Brunner}},\ }\href {\doibase 10.1103/PhysRevA.99.022317} {\bibfield  {journal} {\bibinfo  {journal} {Phys. Rev. A}\ }\textbf {\bibinfo {volume} {99}},\ \bibinfo {pages} {022317} (\bibinfo {year} {2019})}\BibitemShut {NoStop}%
\bibitem [{\citenamefont {Halder}\ \emph {et~al.}(2022)\citenamefont {Halder}, \citenamefont {Banerjee}, \citenamefont {Mal},\ and\ \citenamefont {Sen(De)}}]{Halder_PRA_2022}%
  \BibitemOpen
  \bibfield  {author} {\bibinfo {author} {\bibfnamefont {P.}~\bibnamefont {Halder}}, \bibinfo {author} {\bibfnamefont {R.}~\bibnamefont {Banerjee}}, \bibinfo {author} {\bibfnamefont {S.}~\bibnamefont {Mal}}, \ and\ \bibinfo {author} {\bibfnamefont {A.}~\bibnamefont {Sen(De)}},\ }\href {\doibase 10.1103/PhysRevA.106.052413} {\bibfield  {journal} {\bibinfo  {journal} {Phys. Rev. A}\ }\textbf {\bibinfo {volume} {106}},\ \bibinfo {pages} {052413} (\bibinfo {year} {2022})}\BibitemShut {NoStop}%
\bibitem [{\citenamefont {Brown}\ and\ \citenamefont {Colbeck}(2020)}]{Brown_PRL_2020}%
  \BibitemOpen
  \bibfield  {author} {\bibinfo {author} {\bibfnamefont {P.~J.}\ \bibnamefont {Brown}}\ and\ \bibinfo {author} {\bibfnamefont {R.}~\bibnamefont {Colbeck}},\ }\href {\doibase 10.1103/PhysRevLett.125.090401} {\bibfield  {journal} {\bibinfo  {journal} {Phys. Rev. Lett.}\ }\textbf {\bibinfo {volume} {125}},\ \bibinfo {pages} {090401} (\bibinfo {year} {2020})}\BibitemShut {NoStop}%
\bibitem [{\citenamefont {Pandit}\ \emph {et~al.}(2022)\citenamefont {Pandit}, \citenamefont {Srivastava},\ and\ \citenamefont {Sen}}]{Pandit_PRA_2022}%
  \BibitemOpen
  \bibfield  {author} {\bibinfo {author} {\bibfnamefont {M.}~\bibnamefont {Pandit}}, \bibinfo {author} {\bibfnamefont {C.}~\bibnamefont {Srivastava}}, \ and\ \bibinfo {author} {\bibfnamefont {U.}~\bibnamefont {Sen}},\ }\href {\doibase 10.1103/PhysRevA.106.032419} {\bibfield  {journal} {\bibinfo  {journal} {Physical Review A}\ }\textbf {\bibinfo {volume} {106}},\ \bibinfo {pages} {032419} (\bibinfo {year} {2022})}\BibitemShut {NoStop}%
\bibitem [{\citenamefont {Srivastava}\ \emph {et~al.}(2022{\natexlab{a}})\citenamefont {Srivastava}, \citenamefont {Pandit},\ and\ \citenamefont {Sen}}]{Srivastava_PRA_2022}%
  \BibitemOpen
  \bibfield  {author} {\bibinfo {author} {\bibfnamefont {C.}~\bibnamefont {Srivastava}}, \bibinfo {author} {\bibfnamefont {M.}~\bibnamefont {Pandit}}, \ and\ \bibinfo {author} {\bibfnamefont {U.}~\bibnamefont {Sen}},\ }\href {\doibase 10.1103/PhysRevA.105.062413} {\bibfield  {journal} {\bibinfo  {journal} {Physical Review A}\ }\textbf {\bibinfo {volume} {105}},\ \bibinfo {pages} {062413} (\bibinfo {year} {2022}{\natexlab{a}})}\BibitemShut {NoStop}%
\bibitem [{\citenamefont {Srivastava}\ \emph {et~al.}(2021)\citenamefont {Srivastava}, \citenamefont {Mal}, \citenamefont {Sen(De)},\ and\ \citenamefont {Sen}}]{Srivastava_PRA_2021}%
  \BibitemOpen
  \bibfield  {author} {\bibinfo {author} {\bibfnamefont {C.}~\bibnamefont {Srivastava}}, \bibinfo {author} {\bibfnamefont {S.}~\bibnamefont {Mal}}, \bibinfo {author} {\bibfnamefont {A.}~\bibnamefont {Sen(De)}}, \ and\ \bibinfo {author} {\bibfnamefont {U.}~\bibnamefont {Sen}},\ }\href {\doibase 10.1103/PhysRevA.103.032408} {\bibfield  {journal} {\bibinfo  {journal} {Physical Review A}\ }\textbf {\bibinfo {volume} {103}},\ \bibinfo {pages} {032408} (\bibinfo {year} {2021})}\BibitemShut {NoStop}%
\bibitem [{\citenamefont {Maity}\ \emph {et~al.}(2020)\citenamefont {Maity}, \citenamefont {Das}, \citenamefont {Ghosal}, \citenamefont {Roy},\ and\ \citenamefont {Majumdar}}]{Maity_PRA_2020}%
  \BibitemOpen
  \bibfield  {author} {\bibinfo {author} {\bibfnamefont {A.~G.}\ \bibnamefont {Maity}}, \bibinfo {author} {\bibfnamefont {D.}~\bibnamefont {Das}}, \bibinfo {author} {\bibfnamefont {A.}~\bibnamefont {Ghosal}}, \bibinfo {author} {\bibfnamefont {A.}~\bibnamefont {Roy}}, \ and\ \bibinfo {author} {\bibfnamefont {A.~S.}\ \bibnamefont {Majumdar}},\ }\href {\doibase 10.1103/PhysRevA.101.042340} {\bibfield  {journal} {\bibinfo  {journal} {Phys. Rev. A}\ }\textbf {\bibinfo {volume} {101}},\ \bibinfo {pages} {042340} (\bibinfo {year} {2020})}\BibitemShut {NoStop}%
\bibitem [{\citenamefont {Gupta}\ \emph {et~al.}(2021)\citenamefont {Gupta}, \citenamefont {Maity}, \citenamefont {Das}, \citenamefont {Roy},\ and\ \citenamefont {Majumdar}}]{Gupta_PRA_2021}%
  \BibitemOpen
  \bibfield  {author} {\bibinfo {author} {\bibfnamefont {S.}~\bibnamefont {Gupta}}, \bibinfo {author} {\bibfnamefont {A.~G.}\ \bibnamefont {Maity}}, \bibinfo {author} {\bibfnamefont {D.}~\bibnamefont {Das}}, \bibinfo {author} {\bibfnamefont {A.}~\bibnamefont {Roy}}, \ and\ \bibinfo {author} {\bibfnamefont {A.~S.}\ \bibnamefont {Majumdar}},\ }\href {\doibase 10.1103/PhysRevA.103.022421} {\bibfield  {journal} {\bibinfo  {journal} {Phys. Rev. A}\ }\textbf {\bibinfo {volume} {103}},\ \bibinfo {pages} {022421} (\bibinfo {year} {2021})}\BibitemShut {NoStop}%
\bibitem [{\citenamefont {Srivastava}\ \emph {et~al.}(2022{\natexlab{b}})\citenamefont {Srivastava}, \citenamefont {Pandit},\ and\ \citenamefont {Sen}}]{Srivastava_arXiv_2022}%
  \BibitemOpen
  \bibfield  {author} {\bibinfo {author} {\bibfnamefont {C.}~\bibnamefont {Srivastava}}, \bibinfo {author} {\bibfnamefont {M.}~\bibnamefont {Pandit}}, \ and\ \bibinfo {author} {\bibfnamefont {U.}~\bibnamefont {Sen}},\ }\href {https://arxiv.org/abs/2205.02695} {\bibfield  {journal} {\bibinfo  {journal} {arXiv:2205.02695}\ } (\bibinfo {year} {2022}{\natexlab{b}})}\BibitemShut {NoStop}%
\bibitem [{\citenamefont {Srivastava}\ \emph {et~al.}(2022{\natexlab{c}})\citenamefont {Srivastava}, \citenamefont {Pandit},\ and\ \citenamefont {Sen}}]{Srivastava_arXiv_2022_2}%
  \BibitemOpen
  \bibfield  {author} {\bibinfo {author} {\bibfnamefont {C.}~\bibnamefont {Srivastava}}, \bibinfo {author} {\bibfnamefont {M.}~\bibnamefont {Pandit}}, \ and\ \bibinfo {author} {\bibfnamefont {U.}~\bibnamefont {Sen}},\ }\href {https://arxiv.org/abs/2208.08435} {\bibfield  {journal} {\bibinfo  {journal} {arXiv:2208.08435}\ } (\bibinfo {year} {2022}{\natexlab{c}})}\BibitemShut {NoStop}%
\bibitem [{\citenamefont {Mohan}\ \emph {et~al.}(2019)\citenamefont {Mohan}, \citenamefont {Tavakoli},\ and\ \citenamefont {Brunner}}]{Mohan_NJP_2019}%
  \BibitemOpen
  \bibfield  {author} {\bibinfo {author} {\bibfnamefont {K.}~\bibnamefont {Mohan}}, \bibinfo {author} {\bibfnamefont {A.}~\bibnamefont {Tavakoli}}, \ and\ \bibinfo {author} {\bibfnamefont {N.}~\bibnamefont {Brunner}},\ }\href {\doibase 10.1088/1367-2630/ab3773} {\bibfield  {journal} {\bibinfo  {journal} {New Journal of Physics}\ }\textbf {\bibinfo {volume} {21}},\ \bibinfo {pages} {083034} (\bibinfo {year} {2019})}\BibitemShut {NoStop}%
\bibitem [{\citenamefont {Miklin}\ \emph {et~al.}(2020)\citenamefont {Miklin}, \citenamefont {Borka\l{}a},\ and\ \citenamefont {Paw\l{}owski}}]{Miklin_PRR_2020}%
  \BibitemOpen
  \bibfield  {author} {\bibinfo {author} {\bibfnamefont {N.}~\bibnamefont {Miklin}}, \bibinfo {author} {\bibfnamefont {J.~J.}\ \bibnamefont {Borka\l{}a}}, \ and\ \bibinfo {author} {\bibfnamefont {M.}~\bibnamefont {Paw\l{}owski}},\ }\href {\doibase 10.1103/PhysRevResearch.2.033014} {\bibfield  {journal} {\bibinfo  {journal} {Phys. Rev. Res.}\ }\textbf {\bibinfo {volume} {2}},\ \bibinfo {pages} {033014} (\bibinfo {year} {2020})}\BibitemShut {NoStop}%
\bibitem [{\citenamefont {Curchod}\ \emph {et~al.}(2017)\citenamefont {Curchod}, \citenamefont {Johansson}, \citenamefont {Augusiak}, \citenamefont {Hoban}, \citenamefont {Wittek},\ and\ \citenamefont {Ac\'{\i}n}}]{Curchod_PRA_2017}%
  \BibitemOpen
  \bibfield  {author} {\bibinfo {author} {\bibfnamefont {F.~J.}\ \bibnamefont {Curchod}}, \bibinfo {author} {\bibfnamefont {M.}~\bibnamefont {Johansson}}, \bibinfo {author} {\bibfnamefont {R.}~\bibnamefont {Augusiak}}, \bibinfo {author} {\bibfnamefont {M.~J.}\ \bibnamefont {Hoban}}, \bibinfo {author} {\bibfnamefont {P.}~\bibnamefont {Wittek}}, \ and\ \bibinfo {author} {\bibfnamefont {A.}~\bibnamefont {Ac\'{\i}n}},\ }\href {\doibase 10.1103/PhysRevA.95.020102} {\bibfield  {journal} {\bibinfo  {journal} {Phys. Rev. A}\ }\textbf {\bibinfo {volume} {95}},\ \bibinfo {pages} {020102} (\bibinfo {year} {2017})}\BibitemShut {NoStop}%
\bibitem [{\citenamefont {Roy}\ \emph {et~al.}(2021)\citenamefont {Roy}, \citenamefont {Bera}, \citenamefont {Mal}, \citenamefont {Sen(De)},\ and\ \citenamefont {Sen}}]{Roy_PLA_2021}%
  \BibitemOpen
  \bibfield  {author} {\bibinfo {author} {\bibfnamefont {S.}~\bibnamefont {Roy}}, \bibinfo {author} {\bibfnamefont {A.}~\bibnamefont {Bera}}, \bibinfo {author} {\bibfnamefont {S.}~\bibnamefont {Mal}}, \bibinfo {author} {\bibfnamefont {A.}~\bibnamefont {Sen(De)}}, \ and\ \bibinfo {author} {\bibfnamefont {U.}~\bibnamefont {Sen}},\ }\href {\doibase 10.1016/j.physleta.2021.127143} {\bibfield  {journal} {\bibinfo  {journal} {Physics Letters A}\ }\textbf {\bibinfo {volume} {392}},\ \bibinfo {pages} {127143} (\bibinfo {year} {2021})}\BibitemShut {NoStop}%
\bibitem [{\citenamefont {Das}\ \emph {et~al.}(2023)\citenamefont {Das}, \citenamefont {Halder}, \citenamefont {Banerjee},\ and\ \citenamefont {Sen(De)}}]{Das_PRA_2023}%
  \BibitemOpen
  \bibfield  {author} {\bibinfo {author} {\bibfnamefont {S.}~\bibnamefont {Das}}, \bibinfo {author} {\bibfnamefont {P.}~\bibnamefont {Halder}}, \bibinfo {author} {\bibfnamefont {R.}~\bibnamefont {Banerjee}}, \ and\ \bibinfo {author} {\bibfnamefont {A.}~\bibnamefont {Sen(De)}},\ }\href {\doibase 10.1103/PhysRevA.107.042414} {\bibfield  {journal} {\bibinfo  {journal} {Physical Review A}\ }\textbf {\bibinfo {volume} {107}},\ \bibinfo {pages} {042414} (\bibinfo {year} {2023})}\BibitemShut {NoStop}%
\bibitem [{\citenamefont {Schiavon}\ \emph {et~al.}(2017)\citenamefont {Schiavon}, \citenamefont {Calderaro}, \citenamefont {Pittaluga}, \citenamefont {Vallone},\ and\ \citenamefont {Villoresi}}]{Schiavon_QST_2017}%
  \BibitemOpen
  \bibfield  {author} {\bibinfo {author} {\bibfnamefont {M.}~\bibnamefont {Schiavon}}, \bibinfo {author} {\bibfnamefont {L.}~\bibnamefont {Calderaro}}, \bibinfo {author} {\bibfnamefont {M.}~\bibnamefont {Pittaluga}}, \bibinfo {author} {\bibfnamefont {G.}~\bibnamefont {Vallone}}, \ and\ \bibinfo {author} {\bibfnamefont {P.}~\bibnamefont {Villoresi}},\ }\href {\doibase 10.1088/2058-9565/aa62be} {\bibfield  {journal} {\bibinfo  {journal} {Quantum Science and Technology}\ }\textbf {\bibinfo {volume} {2}},\ \bibinfo {pages} {015010} (\bibinfo {year} {2017})}\BibitemShut {NoStop}%
\bibitem [{\citenamefont {Hu}\ \emph {et~al.}(2018)\citenamefont {Hu}, \citenamefont {Zhou}, \citenamefont {Hu}, \citenamefont {Li}, \citenamefont {Guo},\ and\ \citenamefont {Zhang}}]{Hu_NPJ_2018}%
  \BibitemOpen
  \bibfield  {author} {\bibinfo {author} {\bibfnamefont {M.-J.}\ \bibnamefont {Hu}}, \bibinfo {author} {\bibfnamefont {Z.-Y.}\ \bibnamefont {Zhou}}, \bibinfo {author} {\bibfnamefont {X.-M.}\ \bibnamefont {Hu}}, \bibinfo {author} {\bibfnamefont {C.-F.}\ \bibnamefont {Li}}, \bibinfo {author} {\bibfnamefont {G.-C.}\ \bibnamefont {Guo}}, \ and\ \bibinfo {author} {\bibfnamefont {Y.-S.}\ \bibnamefont {Zhang}},\ }\href {\doibase 10.1038/s41534-018-0115-x} {\bibfield  {journal} {\bibinfo  {journal} {npj Quantum Information}\ }\textbf {\bibinfo {volume} {4}},\ \bibinfo {pages} {63} (\bibinfo {year} {2018})}\BibitemShut {NoStop}%
\bibitem [{\citenamefont {Foletto}\ \emph {et~al.}(2020)\citenamefont {Foletto}, \citenamefont {Calderaro}, \citenamefont {Tavakoli}, \citenamefont {Schiavon}, \citenamefont {Picciariello}, \citenamefont {Cabello}, \citenamefont {Villoresi},\ and\ \citenamefont {Vallone}}]{Foletto_PRAppl_2020}%
  \BibitemOpen
  \bibfield  {author} {\bibinfo {author} {\bibfnamefont {G.}~\bibnamefont {Foletto}}, \bibinfo {author} {\bibfnamefont {L.}~\bibnamefont {Calderaro}}, \bibinfo {author} {\bibfnamefont {A.}~\bibnamefont {Tavakoli}}, \bibinfo {author} {\bibfnamefont {M.}~\bibnamefont {Schiavon}}, \bibinfo {author} {\bibfnamefont {F.}~\bibnamefont {Picciariello}}, \bibinfo {author} {\bibfnamefont {A.}~\bibnamefont {Cabello}}, \bibinfo {author} {\bibfnamefont {P.}~\bibnamefont {Villoresi}}, \ and\ \bibinfo {author} {\bibfnamefont {G.}~\bibnamefont {Vallone}},\ }\href {\doibase 10.1103/PhysRevApplied.13.044008} {\bibfield  {journal} {\bibinfo  {journal} {Phys. Rev. Appl.}\ }\textbf {\bibinfo {volume} {13}},\ \bibinfo {pages} {044008} (\bibinfo {year} {2020})}\BibitemShut {NoStop}%
\bibitem [{\citenamefont {Choi}\ \emph {et~al.}(2020)\citenamefont {Choi}, \citenamefont {Hong}, \citenamefont {Pramanik}, \citenamefont {Lim}, \citenamefont {Kim}, \citenamefont {Jung}, \citenamefont {Han}, \citenamefont {Moon},\ and\ \citenamefont {Cho}}]{Choi_Optica_2020}%
  \BibitemOpen
  \bibfield  {author} {\bibinfo {author} {\bibfnamefont {Y.-H.}\ \bibnamefont {Choi}}, \bibinfo {author} {\bibfnamefont {S.}~\bibnamefont {Hong}}, \bibinfo {author} {\bibfnamefont {T.}~\bibnamefont {Pramanik}}, \bibinfo {author} {\bibfnamefont {H.-T.}\ \bibnamefont {Lim}}, \bibinfo {author} {\bibfnamefont {Y.-S.}\ \bibnamefont {Kim}}, \bibinfo {author} {\bibfnamefont {H.}~\bibnamefont {Jung}}, \bibinfo {author} {\bibfnamefont {S.-W.}\ \bibnamefont {Han}}, \bibinfo {author} {\bibfnamefont {S.}~\bibnamefont {Moon}}, \ and\ \bibinfo {author} {\bibfnamefont {Y.-W.}\ \bibnamefont {Cho}},\ }\href {\doibase 10.1364/OPTICA.394667} {\bibfield  {journal} {\bibinfo  {journal} {Optica}\ }\textbf {\bibinfo {volume} {7}},\ \bibinfo {pages} {675} (\bibinfo {year} {2020})}\BibitemShut {NoStop}%
\bibitem [{\citenamefont {Feng}\ \emph {et~al.}(2020)\citenamefont {Feng}, \citenamefont {Ren}, \citenamefont {Tian}, \citenamefont {Luo}, \citenamefont {Shi}, \citenamefont {Chen},\ and\ \citenamefont {Zhou}}]{Feng_PRA_2020}%
  \BibitemOpen
  \bibfield  {author} {\bibinfo {author} {\bibfnamefont {T.}~\bibnamefont {Feng}}, \bibinfo {author} {\bibfnamefont {C.}~\bibnamefont {Ren}}, \bibinfo {author} {\bibfnamefont {Y.}~\bibnamefont {Tian}}, \bibinfo {author} {\bibfnamefont {M.}~\bibnamefont {Luo}}, \bibinfo {author} {\bibfnamefont {H.}~\bibnamefont {Shi}}, \bibinfo {author} {\bibfnamefont {J.}~\bibnamefont {Chen}}, \ and\ \bibinfo {author} {\bibfnamefont {X.}~\bibnamefont {Zhou}},\ }\href {\doibase 10.1103/PhysRevA.102.032220} {\bibfield  {journal} {\bibinfo  {journal} {Phys. Rev. A}\ }\textbf {\bibinfo {volume} {102}},\ \bibinfo {pages} {032220} (\bibinfo {year} {2020})}\BibitemShut {NoStop}%
\bibitem [{\citenamefont {Das}\ and\ \citenamefont {Arvind}(2014)}]{Das_PRA_2014}%
  \BibitemOpen
  \bibfield  {author} {\bibinfo {author} {\bibfnamefont {D.}~\bibnamefont {Das}}\ and\ \bibinfo {author} {\bibnamefont {Arvind}},\ }\href {\doibase 10.1103/PhysRevA.89.062121} {\bibfield  {journal} {\bibinfo  {journal} {Phys. Rev. A}\ }\textbf {\bibinfo {volume} {89}},\ \bibinfo {pages} {062121} (\bibinfo {year} {2014})}\BibitemShut {NoStop}%
\bibitem [{\citenamefont {Das}\ and\ \citenamefont {Arvind}(2017)}]{Das_JPA_2017}%
  \BibitemOpen
  \bibfield  {author} {\bibinfo {author} {\bibfnamefont {D.}~\bibnamefont {Das}}\ and\ \bibinfo {author} {\bibnamefont {Arvind}},\ }\href {\doibase 10.1088/1751-8121/aa608f} {\bibfield  {journal} {\bibinfo  {journal} {Journal of Physics A: Mathematical and Theoretical}\ }\textbf {\bibinfo {volume} {50}},\ \bibinfo {pages} {145307} (\bibinfo {year} {2017})}\BibitemShut {NoStop}%
\bibitem [{\citenamefont {Braunstein}\ and\ \citenamefont {Kimble}(1998)}]{Braunstein_PRL_1998}%
  \BibitemOpen
  \bibfield  {author} {\bibinfo {author} {\bibfnamefont {S.~L.}\ \bibnamefont {Braunstein}}\ and\ \bibinfo {author} {\bibfnamefont {H.~J.}\ \bibnamefont {Kimble}},\ }\href {\doibase 10.1103/PhysRevLett.80.869} {\bibfield  {journal} {\bibinfo  {journal} {Phys. Rev. Lett.}\ }\textbf {\bibinfo {volume} {80}},\ \bibinfo {pages} {869} (\bibinfo {year} {1998})}\BibitemShut {NoStop}%
\bibitem [{\citenamefont {Braunstein}\ and\ \citenamefont {Kimble}(2000)}]{Braunstein_PRA_2000}%
  \BibitemOpen
  \bibfield  {author} {\bibinfo {author} {\bibfnamefont {S.~L.}\ \bibnamefont {Braunstein}}\ and\ \bibinfo {author} {\bibfnamefont {H.~J.}\ \bibnamefont {Kimble}},\ }\href {\doibase 10.1103/PhysRevA.61.042302} {\bibfield  {journal} {\bibinfo  {journal} {Phys. Rev. A}\ }\textbf {\bibinfo {volume} {61}},\ \bibinfo {pages} {042302} (\bibinfo {year} {2000})}\BibitemShut {NoStop}%
\bibitem [{\citenamefont {Grosshans}\ \emph {et~al.}(2003)\citenamefont {Grosshans}, \citenamefont {Van~Assche}, \citenamefont {Wenger}, \citenamefont {Brouri}, \citenamefont {Cerf},\ and\ \citenamefont {Grangier}}]{Grosshans_Nature_2003}%
  \BibitemOpen
  \bibfield  {author} {\bibinfo {author} {\bibfnamefont {F.}~\bibnamefont {Grosshans}}, \bibinfo {author} {\bibfnamefont {G.}~\bibnamefont {Van~Assche}}, \bibinfo {author} {\bibfnamefont {J.}~\bibnamefont {Wenger}}, \bibinfo {author} {\bibfnamefont {R.}~\bibnamefont {Brouri}}, \bibinfo {author} {\bibfnamefont {N.~J.}\ \bibnamefont {Cerf}}, \ and\ \bibinfo {author} {\bibfnamefont {P.}~\bibnamefont {Grangier}},\ }\href {\doibase 10.1038/nature01289} {\bibfield  {journal} {\bibinfo  {journal} {Nature}\ }\textbf {\bibinfo {volume} {421}},\ \bibinfo {pages} {238} (\bibinfo {year} {2003})}\BibitemShut {NoStop}%
\bibitem [{\citenamefont {Curty}\ \emph {et~al.}(2004)\citenamefont {Curty}, \citenamefont {Lewenstein},\ and\ \citenamefont {L\"utkenhaus}}]{Curty_PRL_2004}%
  \BibitemOpen
  \bibfield  {author} {\bibinfo {author} {\bibfnamefont {M.}~\bibnamefont {Curty}}, \bibinfo {author} {\bibfnamefont {M.}~\bibnamefont {Lewenstein}}, \ and\ \bibinfo {author} {\bibfnamefont {N.}~\bibnamefont {L\"utkenhaus}},\ }\href {\doibase 10.1103/PhysRevLett.92.217903} {\bibfield  {journal} {\bibinfo  {journal} {Phys. Rev. Lett.}\ }\textbf {\bibinfo {volume} {92}},\ \bibinfo {pages} {217903} (\bibinfo {year} {2004})}\BibitemShut {NoStop}%
\bibitem [{\citenamefont {Derka}\ \emph {et~al.}(1998)\citenamefont {Derka}, \citenamefont {Buz\ifmmode~\breve{}\else \u{}\fi{}ek},\ and\ \citenamefont {Ekert}}]{Derka_PRL_1998}%
  \BibitemOpen
  \bibfield  {author} {\bibinfo {author} {\bibfnamefont {R.}~\bibnamefont {Derka}}, \bibinfo {author} {\bibfnamefont {V.}~\bibnamefont {Buz\ifmmode~\breve{}\else \u{}\fi{}ek}}, \ and\ \bibinfo {author} {\bibfnamefont {A.~K.}\ \bibnamefont {Ekert}},\ }\href {\doibase 10.1103/PhysRevLett.80.1571} {\bibfield  {journal} {\bibinfo  {journal} {Phys. Rev. Lett.}\ }\textbf {\bibinfo {volume} {80}},\ \bibinfo {pages} {1571} (\bibinfo {year} {1998})}\BibitemShut {NoStop}%
\bibitem [{\citenamefont {Shang}\ \emph {et~al.}(2018)\citenamefont {Shang}, \citenamefont {Asadian}, \citenamefont {Zhu},\ and\ \citenamefont {G\"uhne}}]{Shang_PRA_2018}%
  \BibitemOpen
  \bibfield  {author} {\bibinfo {author} {\bibfnamefont {J.}~\bibnamefont {Shang}}, \bibinfo {author} {\bibfnamefont {A.}~\bibnamefont {Asadian}}, \bibinfo {author} {\bibfnamefont {H.}~\bibnamefont {Zhu}}, \ and\ \bibinfo {author} {\bibfnamefont {O.}~\bibnamefont {G\"uhne}},\ }\href {\doibase 10.1103/PhysRevA.98.022309} {\bibfield  {journal} {\bibinfo  {journal} {Phys. Rev. A}\ }\textbf {\bibinfo {volume} {98}},\ \bibinfo {pages} {022309} (\bibinfo {year} {2018})}\BibitemShut {NoStop}%
\bibitem [{\citenamefont {Dieks}(1988)}]{Dieks_PLA_1988}%
  \BibitemOpen
  \bibfield  {author} {\bibinfo {author} {\bibfnamefont {D.}~\bibnamefont {Dieks}},\ }\href {\doibase 10.1016/0375-9601(88)90840-7} {\bibfield  {journal} {\bibinfo  {journal} {Physics Letters A}\ }\textbf {\bibinfo {volume} {126}},\ \bibinfo {pages} {303} (\bibinfo {year} {1988})}\BibitemShut {NoStop}%
\bibitem [{\citenamefont {Peres}(1988)}]{Peres_PLA_1988}%
  \BibitemOpen
  \bibfield  {author} {\bibinfo {author} {\bibfnamefont {A.}~\bibnamefont {Peres}},\ }\href {\doibase 10.1016/0375-9601(88)91034-1} {\bibfield  {journal} {\bibinfo  {journal} {Physics Letters A}\ }\textbf {\bibinfo {volume} {128}},\ \bibinfo {pages} {19} (\bibinfo {year} {1988})}\BibitemShut {NoStop}%
\bibitem [{\citenamefont {V\'ertesi}\ and\ \citenamefont {Bene}(2010)}]{Vertesi_PRA_2010}%
  \BibitemOpen
  \bibfield  {author} {\bibinfo {author} {\bibfnamefont {T.}~\bibnamefont {V\'ertesi}}\ and\ \bibinfo {author} {\bibfnamefont {E.}~\bibnamefont {Bene}},\ }\href {\doibase 10.1103/PhysRevA.82.062115} {\bibfield  {journal} {\bibinfo  {journal} {Phys. Rev. A}\ }\textbf {\bibinfo {volume} {82}},\ \bibinfo {pages} {062115} (\bibinfo {year} {2010})}\BibitemShut {NoStop}%
\bibitem [{\citenamefont {G\'omez}\ \emph {et~al.}(2018)\citenamefont {G\'omez}, \citenamefont {Mattar}, \citenamefont {G\'omez}, \citenamefont {Cavalcanti}, \citenamefont {Far\'{\i}as}, \citenamefont {Ac\'{\i}n},\ and\ \citenamefont {Lima}}]{Gomez_PRA_2018}%
  \BibitemOpen
  \bibfield  {author} {\bibinfo {author} {\bibfnamefont {S.}~\bibnamefont {G\'omez}}, \bibinfo {author} {\bibfnamefont {A.}~\bibnamefont {Mattar}}, \bibinfo {author} {\bibfnamefont {E.~S.}\ \bibnamefont {G\'omez}}, \bibinfo {author} {\bibfnamefont {D.}~\bibnamefont {Cavalcanti}}, \bibinfo {author} {\bibfnamefont {O.~J.}\ \bibnamefont {Far\'{\i}as}}, \bibinfo {author} {\bibfnamefont {A.}~\bibnamefont {Ac\'{\i}n}}, \ and\ \bibinfo {author} {\bibfnamefont {G.}~\bibnamefont {Lima}},\ }\href {\doibase 10.1103/PhysRevA.97.040102} {\bibfield  {journal} {\bibinfo  {journal} {Phys. Rev. A}\ }\textbf {\bibinfo {volume} {97}},\ \bibinfo {pages} {040102} (\bibinfo {year} {2018})}\BibitemShut {NoStop}%
\bibitem [{\citenamefont {Halder}\ \emph {et~al.}(2021)\citenamefont {Halder}, \citenamefont {Mal},\ and\ \citenamefont {Sen(De)}}]{Halder_PRA_2021}%
  \BibitemOpen
  \bibfield  {author} {\bibinfo {author} {\bibfnamefont {P.}~\bibnamefont {Halder}}, \bibinfo {author} {\bibfnamefont {S.}~\bibnamefont {Mal}}, \ and\ \bibinfo {author} {\bibfnamefont {A.}~\bibnamefont {Sen(De)}},\ }\href {\doibase 10.1103/PhysRevA.104.062412} {\bibfield  {journal} {\bibinfo  {journal} {Phys. Rev. A}\ }\textbf {\bibinfo {volume} {104}},\ \bibinfo {pages} {062412} (\bibinfo {year} {2021})}\BibitemShut {NoStop}%
\bibitem [{\citenamefont {Mondal}\ \emph {et~al.}(2023)\citenamefont {Mondal}, \citenamefont {Halder},\ and\ \citenamefont {De}}]{Mondal_arXiv_2023}%
  \BibitemOpen
  \bibfield  {author} {\bibinfo {author} {\bibfnamefont {S.}~\bibnamefont {Mondal}}, \bibinfo {author} {\bibfnamefont {P.}~\bibnamefont {Halder}}, \ and\ \bibinfo {author} {\bibfnamefont {A.~S.}\ \bibnamefont {De}},\ }\href {https://arxiv.org/abs/2308.10975} {\bibfield  {journal} {\bibinfo  {journal} {arXiv:2308.10975}\ } (\bibinfo {year} {2023})}\BibitemShut {NoStop}%
\bibitem [{\citenamefont {Nielsen}\ and\ \citenamefont {Chuang}(2000)}]{nielsenchuang}%
  \BibitemOpen
  \bibfield  {author} {\bibinfo {author} {\bibfnamefont {M.}~\bibnamefont {Nielsen}}\ and\ \bibinfo {author} {\bibfnamefont {I.}~\bibnamefont {Chuang}},\ }\href {\doibase https://doi.org/10.1017/CBO9780511976667} {\emph {\bibinfo {title} {Quantum Computation and Quantum Information}}}\ (\bibinfo  {publisher} {Cambridge University Press},\ \bibinfo {year} {2000})\BibitemShut {NoStop}%
\bibitem [{\citenamefont {van Loock}\ and\ \citenamefont {Braunstein}(2001)}]{van-Loock_PRL_2001}%
  \BibitemOpen
  \bibfield  {author} {\bibinfo {author} {\bibfnamefont {P.}~\bibnamefont {van Loock}}\ and\ \bibinfo {author} {\bibfnamefont {S.~L.}\ \bibnamefont {Braunstein}},\ }\href {\doibase 10.1103/PhysRevLett.87.247901} {\bibfield  {journal} {\bibinfo  {journal} {Phys. Rev. Lett.}\ }\textbf {\bibinfo {volume} {87}},\ \bibinfo {pages} {247901} (\bibinfo {year} {2001})}\BibitemShut {NoStop}%
\bibitem [{\citenamefont {Einstein}\ \emph {et~al.}(1935)\citenamefont {Einstein}, \citenamefont {Podolsky},\ and\ \citenamefont {Rosen}}]{Einstein_PR_1935}%
  \BibitemOpen
  \bibfield  {author} {\bibinfo {author} {\bibfnamefont {A.}~\bibnamefont {Einstein}}, \bibinfo {author} {\bibfnamefont {B.}~\bibnamefont {Podolsky}}, \ and\ \bibinfo {author} {\bibfnamefont {N.}~\bibnamefont {Rosen}},\ }\href {\doibase 10.1103/PhysRev.47.777} {\bibfield  {journal} {\bibinfo  {journal} {Phys. Rev.}\ }\textbf {\bibinfo {volume} {47}},\ \bibinfo {pages} {777} (\bibinfo {year} {1935})}\BibitemShut {NoStop}%
\bibitem [{\citenamefont {Duan}\ \emph {et~al.}(2000)\citenamefont {Duan}, \citenamefont {Giedke}, \citenamefont {Cirac},\ and\ \citenamefont {Zoller}}]{Duan_PRL_2000}%
  \BibitemOpen
  \bibfield  {author} {\bibinfo {author} {\bibfnamefont {L.-M.}\ \bibnamefont {Duan}}, \bibinfo {author} {\bibfnamefont {G.}~\bibnamefont {Giedke}}, \bibinfo {author} {\bibfnamefont {J.~I.}\ \bibnamefont {Cirac}}, \ and\ \bibinfo {author} {\bibfnamefont {P.}~\bibnamefont {Zoller}},\ }\href {\doibase 10.1103/PhysRevLett.84.2722} {\bibfield  {journal} {\bibinfo  {journal} {Phys. Rev. Lett.}\ }\textbf {\bibinfo {volume} {84}},\ \bibinfo {pages} {2722} (\bibinfo {year} {2000})}\BibitemShut {NoStop}%
\bibitem [{\citenamefont {Das}\ \emph {et~al.}(2024)\citenamefont {Das}, \citenamefont {Gupta}, \citenamefont {Dhar},\ and\ \citenamefont {Sen(De)}}]{Das_PRA_2024}%
  \BibitemOpen
  \bibfield  {author} {\bibinfo {author} {\bibfnamefont {S.}~\bibnamefont {Das}}, \bibinfo {author} {\bibfnamefont {R.}~\bibnamefont {Gupta}}, \bibinfo {author} {\bibfnamefont {H.~S.}\ \bibnamefont {Dhar}}, \ and\ \bibinfo {author} {\bibfnamefont {A.}~\bibnamefont {Sen(De)}},\ }\href {\doibase 10.1103/PhysRevA.110.012410} {\bibfield  {journal} {\bibinfo  {journal} {Phys. Rev. A}\ }\textbf {\bibinfo {volume} {110}},\ \bibinfo {pages} {012410} (\bibinfo {year} {2024})}\BibitemShut {NoStop}%
\bibitem [{\citenamefont {Cheng}\ \emph {et~al.}(2022)\citenamefont {Cheng}, \citenamefont {Liu}, \citenamefont {Baker},\ and\ \citenamefont {Hall}}]{Cheng_PRA_2022}%
  \BibitemOpen
  \bibfield  {author} {\bibinfo {author} {\bibfnamefont {S.}~\bibnamefont {Cheng}}, \bibinfo {author} {\bibfnamefont {L.}~\bibnamefont {Liu}}, \bibinfo {author} {\bibfnamefont {T.~J.}\ \bibnamefont {Baker}}, \ and\ \bibinfo {author} {\bibfnamefont {M.~J.~W.}\ \bibnamefont {Hall}},\ }\href {\doibase 10.1103/PhysRevA.105.022411} {\bibfield  {journal} {\bibinfo  {journal} {Phys. Rev. A}\ }\textbf {\bibinfo {volume} {105}},\ \bibinfo {pages} {022411} (\bibinfo {year} {2022})}\BibitemShut {NoStop}%
\bibitem [{\citenamefont {Kumari}\ and\ \citenamefont {Pan}(2023)}]{Kumari_PRA_2023}%
  \BibitemOpen
  \bibfield  {author} {\bibinfo {author} {\bibfnamefont {A.}~\bibnamefont {Kumari}}\ and\ \bibinfo {author} {\bibfnamefont {A.~K.}\ \bibnamefont {Pan}},\ }\href {\doibase 10.1103/PhysRevA.107.012615} {\bibfield  {journal} {\bibinfo  {journal} {Phys. Rev. A}\ }\textbf {\bibinfo {volume} {107}},\ \bibinfo {pages} {012615} (\bibinfo {year} {2023})}\BibitemShut {NoStop}%
\bibitem [{\citenamefont {Foletto}\ \emph {et~al.}(2021)\citenamefont {Foletto}, \citenamefont {Padovan}, \citenamefont {Avesani}, \citenamefont {Tebyanian}, \citenamefont {Villoresi},\ and\ \citenamefont {Vallone}}]{Foletto_PRA_2021}%
  \BibitemOpen
  \bibfield  {author} {\bibinfo {author} {\bibfnamefont {G.}~\bibnamefont {Foletto}}, \bibinfo {author} {\bibfnamefont {M.}~\bibnamefont {Padovan}}, \bibinfo {author} {\bibfnamefont {M.}~\bibnamefont {Avesani}}, \bibinfo {author} {\bibfnamefont {H.}~\bibnamefont {Tebyanian}}, \bibinfo {author} {\bibfnamefont {P.}~\bibnamefont {Villoresi}}, \ and\ \bibinfo {author} {\bibfnamefont {G.}~\bibnamefont {Vallone}},\ }\href {\doibase 10.1103/PhysRevA.103.062206} {\bibfield  {journal} {\bibinfo  {journal} {Phys. Rev. A}\ }\textbf {\bibinfo {volume} {103}},\ \bibinfo {pages} {062206} (\bibinfo {year} {2021})}\BibitemShut {NoStop}%
\bibitem [{\citenamefont {Zhang}\ \emph {et~al.}(2015)\citenamefont {Zhang}, \citenamefont {Datta},\ and\ \citenamefont {Walmsley}}]{Zhang_PRL_2015}%
  \BibitemOpen
  \bibfield  {author} {\bibinfo {author} {\bibfnamefont {L.}~\bibnamefont {Zhang}}, \bibinfo {author} {\bibfnamefont {A.}~\bibnamefont {Datta}}, \ and\ \bibinfo {author} {\bibfnamefont {I.~A.}\ \bibnamefont {Walmsley}},\ }\href {\doibase 10.1103/PhysRevLett.114.210801} {\bibfield  {journal} {\bibinfo  {journal} {Phys. Rev. Lett.}\ }\textbf {\bibinfo {volume} {114}},\ \bibinfo {pages} {210801} (\bibinfo {year} {2015})}\BibitemShut {NoStop}%
\bibitem [{\citenamefont {Troupe}\ and\ \citenamefont {Farinholt}(2017)}]{Troupe_arXiv_2017}%
  \BibitemOpen
  \bibfield  {author} {\bibinfo {author} {\bibfnamefont {J.~E.}\ \bibnamefont {Troupe}}\ and\ \bibinfo {author} {\bibfnamefont {J.~M.}\ \bibnamefont {Farinholt}},\ }\href {https://arxiv.org/abs/1702.04836} {\bibfield  {journal} {\bibinfo  {journal} {arXiv:1702.04836}\ } (\bibinfo {year} {2017})}\BibitemShut {NoStop}%
\bibitem [{\citenamefont {Riley}\ \emph {et~al.}(2006)\citenamefont {Riley}, \citenamefont {Hobson},\ and\ \citenamefont {Bence}}]{Riley_Hobson_Bence_2006}%
  \BibitemOpen
  \bibfield  {author} {\bibinfo {author} {\bibfnamefont {K.~F.}\ \bibnamefont {Riley}}, \bibinfo {author} {\bibfnamefont {M.~P.}\ \bibnamefont {Hobson}}, \ and\ \bibinfo {author} {\bibfnamefont {S.~J.}\ \bibnamefont {Bence}},\ }\href@noop {} {\emph {\bibinfo {title} {Mathematical Methods for Physics and Engineering: A Comprehensive Guide}}},\ \bibinfo {edition} {3rd}\ ed.\ (\bibinfo  {publisher} {Cambridge University Press},\ \bibinfo {year} {2006})\BibitemShut {NoStop}%
\end{thebibliography}%
\end{document}